\documentclass[sigconf]{acmart}
\AtBeginDocument{%
}
\copyrightyear{2026}
\acmYear{2026}
\setcopyright{cc}
\setcctype{by}
\acmConference[CHI '26]{Proceedings of the 2026 CHI Conference on Human Factors in Computing Systems}{April 13--17, 2026}{Barcelona, Spain}
\acmBooktitle{Proceedings of the 2026 CHI Conference on Human Factors in Computing Systems (CHI '26), April 13--17, 2026, Barcelona, Spain}
\acmPrice{}
\acmDOI{10.1145/3772318.3790960}
\acmISBN{979-8-4007-2278-3/2026/04}
\usepackage{graphicx}
\usepackage{array}

\usepackage[utf8]{inputenc}
\usepackage[super]{nth}
\usepackage{fancyhdr,amsmath,amsthm,url,enumitem,color,colortbl,etoolbox,multirow,makecell,pifont}
\usepackage{xspace}
\usepackage{booktabs}
\usepackage{subcaption}
\PassOptionsToPackage{margin=1in}{geometry}
\usepackage{geometry}
\usepackage{float} 
\usepackage{xcolor}
\usepackage{hyperref}

\usepackage{amssymb}
\usepackage{rotating}
\newcommand{\complete}{\includegraphics[height=1em]{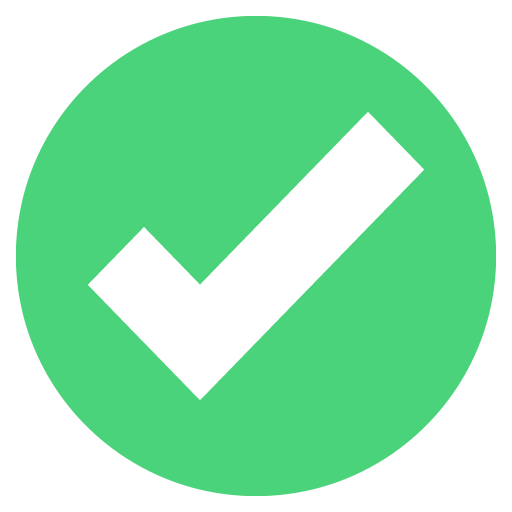}}
\newcommand{\notcomplete}{\includegraphics[height=1em]{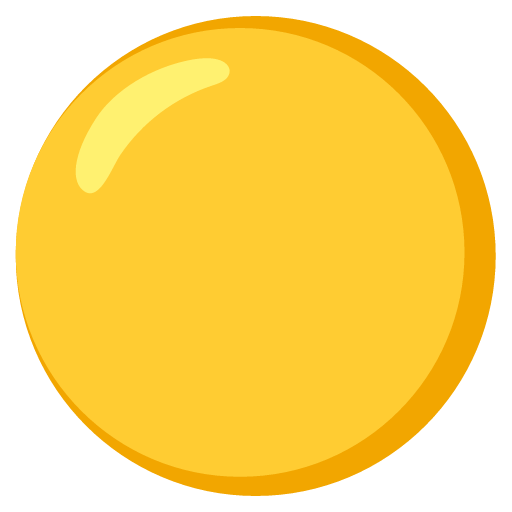}} 
\newcommand{\broken}{\includegraphics[height=1em]{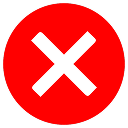}} 
\newcommand{\notapplicable}{\includegraphics[height=1em]{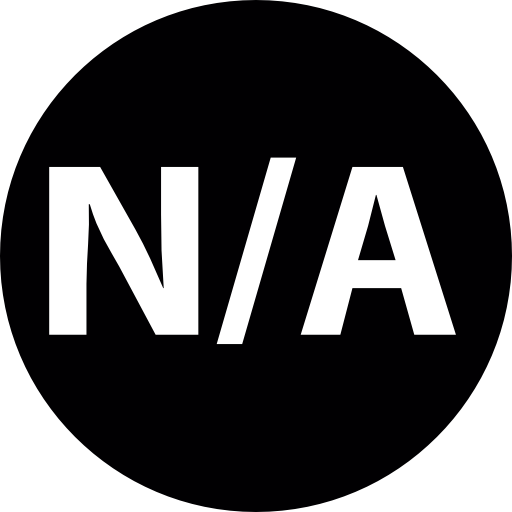}} 
\usepackage{framed}
\definecolor{shadecolor}{gray}{0.95}
\begin{document}
%\begin{CJK*}{UTF8}{gbsn}

%% allowing the author to define a "short title" to be used in page headers.%
\title{Investigating Bystander Privacy in Chinese Smart Home Apps}
%\author{Anonymous Author(s)%

\author{Shijing He}
\email{shijing.he@kcl.ac.uk}
\affiliation{
  \institution{King's College London}
  \city{London}
  \country{United Kingdom}
}

\author{Xuchen Wang}
\email{xuchen.wang@kcl.ac.uk}
\affiliation{
  \institution{King's College London}
  \city{London}
  \country{United Kingdom}
}

\author{Yaxiong Lei}
\email{yl212@st-andrews.ac.uk}
\affiliation{
  \institution{University of St Andrews}
  \city{St Andrews}
  \country{United Kingdom}
}

\author{Chi Zhang}
\email{chi.zhang.9@warwick.ac.uk}
\affiliation{
  \institution{University of Warwick}
  \city{Coventry}
  \country{United Kingdom}
}

\author{Ruba Abu-Salma}
\email{ruba.abu-salma@kcl.ac.uk}
\affiliation{
  \institution{King's College London}
  \city{London}
  \country{United Kingdom}
}

\author{Jose Such}
\email{jose.such@csic.es}
\affiliation{
  \institution{INGENIO (CSIC-Universitat Politècnica de València)}
  \city{Valencia}
  \country{Spain}
}
%% of authors' names for this purpose.%%
\renewcommand{\shortauthors}{He et al.}
\begin{abstract}
Bystander privacy in smart homes has been widely studied in Western contexts, yet it remains underexplored in non-Western countries such as China. In this study, we analyze 49 Chinese smart home apps using a mixed-methods approach, including privacy policy review, UX/UI evaluation, and assessment of Apple App Store privacy labels. While most apps nominally comply with national regulations, we identify significant gaps between written policies and actual implementation. Our traceability analysis highlights inconsistencies in data controls and a lack of transparency in data-sharing practices. Crucially, bystander privacy---particularly for visitors and non-user individuals---is largely absent from both policy documents and interface design. Additionally, discrepancies between privacy labels and actual data practices threaten user trust and undermine informed consent. We provide design recommendations to strengthen bystander protections, improve privacy-oriented UI transparency, and enhance the credibility of privacy labels, supporting the development of inclusive smart home ecosystems in non-Western contexts.
\end{abstract}
\begin{CCSXML}
<ccs2012>
   <concept>
       <concept_id>10002978.10003029.10003032</concept_id>
       <concept_desc>Security and privacy~Social aspects of security and privacy</concept_desc>
       <concept_significance>500</concept_significance>
       </concept>
   <concept>
       <concept_id>10003120.10003121.10011748</concept_id>
       <concept_desc>Human-centered computing~Empirical studies in HCI</concept_desc>
       <concept_significance>300</concept_significance>
       </concept>
 </ccs2012>
\end{CCSXML}

\ccsdesc[500]{Security and privacy~Social aspects of security and privacy}
\ccsdesc[300]{Human-centered computing~Empirical studies in HCI}

\keywords{Smart Homes, Bystanders, Privacy, Privacy Policies, UX/UI Evaluation, Traceability Analysis}


\maketitle

\section{Introduction}
The widespread adoption of smart homes introduces significant privacy and security risks. Existing research---primarily conducted in Western contexts---has examined not only the privacy concerns of primary users~\cite{zeng2017end,mare2020smart,huang2020amazon,abdi2019more,abdi2021privacy,thakkar2022would,barbosa2019if,haney2023user,lau2018alexa,apthorpe2018discovering,he2018rethinking}, but also those of bystanders, including visiting family members, guests, neighbors, and domestic workers~\cite{bernd2020bystanders,yao2019privacy,thakkar2022would,ahmad2020tangible,ahmad2022tangible,albayaydh2022exploring,marky2020idont,alshehri2022exploring,alshehri2023exploring} (see \cite{saqib2025bystander} for a systematic review of smart home bystander privacy concerns). These individuals may be inadvertently affected by smart home devices through the collection, storage, and potential exploitation of their data, unauthorized access to sensitive information, and/or intrusive surveillance by device owners.
%China, as a non-WEIRD country, has experienced a rapid proliferation of smart home technologies, revolutionizing how users interact with their living environments by offering unprecedented levels of convenience, connectivity, and control~\cite{mckinsey}. The high prevalence of these technologies is evidenced by the fact that over 78 million Chinese households used smart home devices in 2023~\cite{chinadata}. These innovations allow individuals to manage various aspects of their homes remotely, including security systems (e.g., smart home cameras, smart door locks/doorbells), entertainment devices (e.g., smart TVs, smart speakers), and household appliances (e.g., smart robot vacuums, smart kitchen appliances), enhancing everyday life.

However, a significant gap remains in understanding how bystander privacy is addressed and designed for in non-Western countries~\cite{despres2024my,saqib2025bystander,abu-salma2025grand}, particularly in the Chinese context, where regulatory frameworks and cultural attitudes toward privacy differ markedly from those in Western settings~\cite{he2025exploring,he2025living,hasegawa2024weird,linxen2021weird}, shaping both corporate practices and consumer expectations. China's collectivist society, which emphasizes national interests and often places less weight on individual privacy~\cite{johnston2009national,li2017cross,zhang2023individualism,he2025exploring,wang2024justifying}, presents unique challenges and opportunities for smart home companies regarding privacy and security, particularly in relation to bystander protections.

Additionally, China's rapid technological development~\cite{mckinsey,chinadata}, combined with a growing market for smart home devices, has outpaced the establishment of comprehensive regulatory frameworks that specifically address bystander privacy~\cite{he2025exploring}. This regulatory lag creates a precarious environment in which the privacy and security of bystanders may be at risk. Given these complexities, it is essential to understand \textbf{how Chinese smart home companies articulate and manage privacy and security concerns, both in their policies and through the UX (user experience) and UI (user interface) design of privacy and security settings affecting bystanders}. These companies operate within a distinctive socio-cultural and legal context that shapes their approach to data protection and bystander privacy. This study aims to address the following research questions (RQs):%

\begin{enumerate}
\item[RQ1:] \emph{How do Chinese smart home apps communicate their privacy and security practices in their privacy policies, particularly with regard to bystanders?}
\item[RQ2:] \emph{How do Chinese smart home apps articulate and manage privacy and security practices through their UX/UI design, particularly with regard to bystanders?}
\item[RQ3:] \emph{What areas of disconnect exist between Chinese smart home companies' actual privacy and security practices in UX/UI design and the practices disclosed in their privacy policies, particularly with regard to bystanders?}
\end{enumerate}%%

To address these RQs, we adopted a mixed-methods approach comprising three components: 1) applying inductive thematic analysis~\cite{braun2006using} to examine the privacy and security practices of Chinese smart home apps in their privacy policies ($n=49$) \textbf{(RQ1)}; 2) conducting cognitive walkthroughs~\cite{blackmon2002cognitive} and heuristic evaluations~\cite{nielsen1990heuristic}, guided by privacy-by-design principles~\cite{cavoukian2009privacy}, to assess the UX/UI practices of these apps \textbf{(RQ2)}; and 3) performing traceability analysis~\cite{anthonysamy2013social,edu2021skillvet} to identify gaps between stated privacy policies and their implementation in app UIs, thereby highlighting potential compliance issues and practical design shortcomings \textbf{(RQ3)}.

\textbf{Contributions.} To our knowledge, this study is the first to comprehensively analyze the app privacy policies and statements issued by Chinese smart home companies, with a particular focus on how these companies articulate and manage privacy and security practices, especially with regard to bystanders. This paper makes three primary contributions. First, it presents a comprehensive traceability analysis of Chinese smart home apps, mapping privacy policy claims against real-world UX/UI privacy practices across 49 apps. Second, it identifies a critical gap in bystander privacy, showing that current policies and interfaces overwhelmingly focus on primary users while neglecting visiting bystanders (e.g., domestic workers) and uninvolved bystanders (e.g., neighbors), raising ethical and regulatory concerns in shared domestic environments. Third, it offers actionable design interventions and regulatory recommendations to enhance bystander-aware privacy protections, improve UI transparency, and strengthen the credibility of app store privacy labels. Collectively, these contributions advance both empirical understanding and practical approaches to inclusive privacy in non-Western smart home contexts.%%
\section{Related Work}
\subsection{Smart Home Bystander Privacy} \label{bystander_privacy} To provide greater convenience and utility, smart home devices collect extensive amounts of data and often continuously monitor the environments in which they are deployed, affecting bystanders---individuals exposed to data collection in environments where they have little or no control over the devices~\cite{yao2019privacy,pierce2022addressing,thakkar2022would}. These devices are used in various contexts, such as monitoring domestic workers, guests, or visitors at the front door~\cite{bernd2020bystanders,albayaydh2022exploring,albayaydh2023examining,slupska2022they,zeng2019understanding,huang2020amazon}. Bystanders often have limited control over the devices’ data practices and may be unaware of their presence or function, lacking knowledge about what data is collected, how it is used, and who has access to it~\cite{yao2019defending,albayaydh2022exploring,h2022monitoring}. Saqib et al.~\cite{saqib2025bystander} classify bystanders based on their relationship to the smart home device owner: visiting bystanders (e.g., visitors and friends~\cite{cobb2021would,alshehri2022exploring,thakkar2022would,geeng2019s,h2022monitoring,windl2023investigating,despres2024my,zhou2024bring}), live-in bystanders (e.g., tenants and family members~\cite{huang2020amazon,thakkar2022would,h2022monitoring,chiang2024more,marky2024decide,shalawadi2024manual,chalhoub2024useful}), and uninvolved bystanders (e.g., neighbors and passersby~\cite{pierce2022addressing,cobb2021would,h2022monitoring,tabassum2023exploring,chiang2024more}). The social relationship between bystanders and device owners strongly shapes privacy concerns: those unfamiliar with the owner, such as temporary visitors, or lacking control over the space, often feel uneasy about surveillance and constant monitoring~\cite{mare2020smart,bernd2022balancing,cobb2021would,marky2020idont,albayaydh2022exploring,liu2024cctv,albayaydh2023examining,wong2023broadening}, whereas closely connected bystanders, such as family or friends, tend to be less concerned about how their data is used~\cite{yao2019privacy,thakkar2022would,zeng2017end}.%%%\textbf{Power imbalances.} Literature highlighted power imbalances can limit bystanders' ability to raise privacy concerns to device owners~\cite{bernd2020bystanders,thakkar2022would,alshehri2023exploring,despres2024my}. For instance, in employer-domestic worker relationships, bystanders (workers) may feel restricted from voicing their privacy concerns due to fears related to job security or social repercussions~\cite{bernd2020bystanders,albayaydh2022exploring,albayaydh2024co,ju2023re}. Further, trust issues extend to device manufacturers, with bystanders worried about data being shared with third parties or exploited for targeted advertising~\cite{huang2020amazon,geeng2019s}.

\subsection{IoT Privacy Policy Analysis} \label{policy_analysis}
%Privacy policy analysis in the context of smart home/IoT devices has emerged as a critical area of research, driven by the proliferation of interconnected technologies and their associated data risks.%

Prior studies have highlighted the complexity and opacity of privacy policies, which often fail to clearly communicate data practices to users~\cite{wu2012effect,mcdonald2008cost,manandhar2022smart,liu2014step,javed2021privacy,kaplan2021lattice,mhaidli2023researchers}. For example, Paul et al. analyzed 94 IoT device privacy policies and identified significant ambiguities concerning data retention and third-party data sharing practices~\cite{paul2018assessing}. Novikova et al. developed an ontology-based framework to assess IoT privacy risks, emphasizing semantic ambiguities and inconsistent policy structures~\cite{novikova2020p2onto}. These findings reinforce broader critiques that privacy policies are overly legalistic and often inaccessible, particularly in technical IoT environments~\cite{mohammadi2019pattern}. Machine learning (ML) and natural language processing (NLP) techniques have been applied to automate IoT privacy policy analysis~\cite{liu2021machine,del2022systematic,guaman2023automated}. For instance, Manandhar et al. used Device Attribute Mapping (DAM) to examine 2,442 smart home device policies, uncovering widespread inconsistencies: 10.57\% of vendors lacked any privacy policy, and 43.52\% failed to explicitly address how device data was handled, despite GDPR requirements~\cite{manandhar2022smart}. However, automated methods face challenges, including limited standard datasets and evolving legal terminology~\cite{del2022systematic,liu2021machine}. Visual and user-centric tools, such as layered policy designs and privacy icons, have also been proposed, though their effectiveness varies with user interpretation and cultural context~\cite{bhattacharjee2020privacy}.

Some studies have specifically examined privacy policies in Chinese contexts. Liu et al. identified notable gaps in Android app compliance across China and Europe: third-party services were responsible for 68\% of data leaks, and while Chinese app stores generally showed better compliance due to regulatory templates, many Google Play apps lacked essential disclosures regarding data handling~\cite{liu2022evaluating}. Lin et al. highlighted inconsistencies between Chinese website policies and actual practices, including long-term cookies and opaque tracking; despite moderate transparency, functional consent mechanisms and oversight remained limited~\cite{lin2022privacy}. However, these studies did not focus on smart home ecosystems or explore the interplay between privacy policies and app UI designs.%

\subsection{Chinese Socio-Legal Context and Regulatory Practices} \label{chinese_contexts}
China's approach to digital privacy is shaped by its unique socio-political environment, in which the state plays a central role in regulating and overseeing digital ecosystems \cite{wang2024justifying,he2025privacy}. The legal framework for digital privacy has expanded in recent years, with laws such as the Personal Information Protection Law (PIPL, 2021)\footnote{\href{http://en.npc.gov.cn.cdurl.cn/2021-12/29/c_694559.htm}{Personal Information Protection Law (PIPL)}}, the Cybersecurity Law (CSL, 2017)\footnote{\href{http://www.npc.gov.cn/zgrdw/npc/xinwen/2016-11/07/content_2001605.htm}{Cybersecurity Law (CSL) of the People's Republic of China}}, and the Data Security Law (DSL, 2021)\footnote{\href{https://www.gov.cn/}{Data Security Law (DSL) of the People's Republic of China}}. These laws reflect a move toward more comprehensive data protection standards aimed at safeguarding personal information and regulating corporate data practices. However, the degree of enforcement and the actual impact of these regulations on organizational behavior remain active areas of inquiry. Prior research indicates that, despite a robust legal framework on paper, inconsistencies in enforcement and ongoing compliance challenges persist \cite{creemers2022china,yao2023overcoming,zhang2024china}.

In addition to legal factors, cultural attitudes toward privacy in China also shape user expectations and corporate practices. Chinese privacy norms are often influenced by collectivist values, where communal harmony and shared benefits may take precedence over individual rights \cite{he2025privacy,he2025living,he2025exploring}. As a result, users may be more accepting of extensive data collection, driven by trust in authorities or perceived communal advantages. Companies, in turn, may design privacy policies and interfaces that reflect these cultural expectations. \cite{he2025privacy} examined how cultural and socio-legal dynamics shape corporate privacy practices, particularly in the design and development of smart home technologies. Understanding these broader contexts is essential for interpreting the privacy strategies of Chinese smart home companies and identifying potential gaps between policy and practice.%%In addition to legal factors, cultural attitudes toward privacy in China also shape user expectations and corporate practices. Chinese privacy norms are often characterized by collectivist values, where communal harmony and shared benefits may take precedence over individual rights~\cite{he2025privacy,he2025living,he2025exploring}. As a result, users may be more accepting of extensive data collection, motivated by trust in authorities or perceived communal advantages. Companies, in turn, may design privacy policies and interfaces that align with these cultural expectations~\cite{he2025privacy}.

%Our study examines how these socio-legal dynamics influence corporate privacy practices, particularly in the design and development of smart home technologies. Understanding these broader contexts is essential for interpreting the privacy strategies of Chinese smart home companies and identifying potential areas of disconnect between policy and practice.%%%

\subsection{Contributions over Previous Works}
Although prior research has extensively documented risks to non-owners in Western contexts---categorizing bystanders by household relationships and highlighting context-dependent concerns (\S\ref{bystander_privacy})---the existing literature has largely overlooked non-Western contexts. Similarly, studies on IoT privacy policies have critiqued their opacity and inconsistencies, often using automated methods focused on legal or technical compliance, particularly in Western jurisdictions (\S\ref{policy_analysis}). While recent scholarship has begun to explore China's evolving privacy landscape (\S\ref{chinese_contexts}), these efforts remain siloed: policy analyses tend to isolate legal interpretation, while UX/UI evaluations are rarely connected to broader legal and regulatory frameworks or sociocultural norms.

To address these gaps, our research is the first to present a comprehensive privacy policy analysis (\S\ref{privacy_policy}), UX/UI evaluation (\S\ref{ux_analyse}), and traceability analysis (\S\ref{traceability_analyse}) specifically tailored to Chinese smart home environments. This approach highlights previously underexplored dimensions of bystander privacy, provides visibility into how bystanders' data is processed, and examines the deletion or opt-out options available for individuals who do not own the devices. Consistent with prior research on privacy policies across various digital technologies and ecosystems, we find that the privacy policies of Chinese smart home apps are excessively lengthy and difficult for typical users to understand (see \S\ref{privacy_policy}), echoing challenges documented in studies of mobile apps more broadly~\cite{mcdonald2008cost,jensen2004privacy,zimmeck2017automated,rao2016expecting}. In other words, smart home apps---being mobile apps themselves---share the same readability and usability issues as other app genres. The technical and legalistic language commonly used in these policies often obscures critical privacy information, hindering informed consent and limiting users' understanding of companies' data practices~\cite{reidenberg2015disagreeable,martin2015privacy}.

Additionally, prior studies have identified ongoing tensions between individual privacy rights and national security or public safety obligations, particularly in jurisdictions enforcing strict data localization and government access requirements~\cite{calzada2022citizens,he2025exploring,he2025privacy}. Our analysis further contributes to understanding this tension within the Chinese context through both narrative and practical assessments, revealing how national security mandates explicitly outlined in privacy policies frequently override individual rights and user agency. This prioritization exacerbates privacy risks, particularly for sensitive or special-category data involving bystanders.

Our study not only corroborates earlier findings on the readability and transparency challenges of privacy policies but also extends this literature by detailing mismatches between privacy policies and UX/UI design, while emphasizing critical gaps in bystander privacy within Chinese smart home ecosystems.%%%To address these gaps, our research is the first to present a comprehensive policy analysis (\S \ref{privacy_policy}), UX/UI (\S \ref{ux_analyse}) and traceability analysis (\S \ref{traceability_analyse}) specifically tailored to Chinese smart home ecosystems, highlighting previously underexplored dimensions of bystander privacy, limited visibility into how bystanders' data is processed, and the absence of deletion or opt-out options for individuals who do not own the device. Consistent with previous research on privacy policies across various digital ecosystems, our study finds that the privacy policies of Chinese smart home apps are excessively lengthy and difficult for typical users to understand (see \S \ref{privacy_policy})
\section{Methodology}
To address our three research questions, we employed a mixed-methods approach organized into three distinct stages. %%In this section, we introduce app collection practices (\S \ref{method_app_collection}), privacy policy analysis (\S \ref{method_privacy_analysis}) to address \textbf{RQ1}, UX/UI analysis (\S \ref{method_ux_analysis}) to address \textbf{RQ2}, and traceability analysis (\S \ref{method_traceablity}) to address \textbf{RQ3}.
%To address \textbf{RQ1}, we first used a qualitative research design to analyze the selected Chinese smart home apps' privacy policies ($n=49$), utilizing inductive thematic analysis~\cite{braun2006using}. Second, we primarily used cognitive walkthroughs~\cite{blackmon2002cognitive} and heuristic evaluations~\cite{nielsen1990heuristic}, and complemented by privacy-by-design principles~\cite{cavoukian2009privacy} to analyze UX/UI practices in the privacy and security sections of these 49 apps \textbf{(RQ2)}. Thirdly, after conducting the first two stages, we used traceability analysis~\cite{anthonysamy2013social,edu2021skillvet} to reveal gaps between what was promised in privacy policies and what was practically implemented in UX/UI design, thus clarifying potential compliance and practical design shortcomings to address \textbf{RQ3}. %The methodology is structured to allow a comprehensive exploration of privacy policies, UX practices, and the socio-cultural context in which these practices are embedded. %
Our study analyzed publicly available apps and did not involve direct interaction with users or access to personal data beyond what is normally collected during app installation. Consequently, approval from our institution’s Research Ethics Office was not required. All evaluations were conducted ethically and in full compliance with each app’s terms of service.%%

\subsection{Collection of Apps} \label{method_app_collection}

We first conducted keyword searches across the Apple App Store in the Mainland China market in January 2025, using keywords such as ``smart/intelligent home,'' ``intelligent/smart life,'' ``smart/intelligent system,'' and ``smart/intelligent device,'' and reviewed the top 200 apps in the categories of Entertainment, Social Networking, Utilities, Weather, Business, Lifestyle, Productivity, Photo \& Video, Food \& Drink, Kids, Education, Health \& Fitness, and Medical Apps. This initial search yielded 122 apps. We excluded apps that had not been updated for over a year ($n=24$) or had fewer than 100 user ratings ($n=31$), and for the remaining 67 apps, we reviewed the developers or service providers, excluding IoT platforms intended for developers (e.g., \href{https://developer.aqara.com/}{Aqara Developer Platform}) and ODM/OEM companies (e.g., \href{https://www.quectel.com/}{Quectel}) that manufacture IoT modules such as Zigbee and NB-IoT ($n=10$). Our final list includes smart home companies founded and based solely in Mainland China, while apps developed by multinational corporations, such as Amazon Alexa, were excluded ($n=6$), and two apps targeting non-Mainland markets (e.g., Hong Kong SAR) were removed.%

\begin{figure*}[h]
\centering
\includegraphics[width=0.8\linewidth]{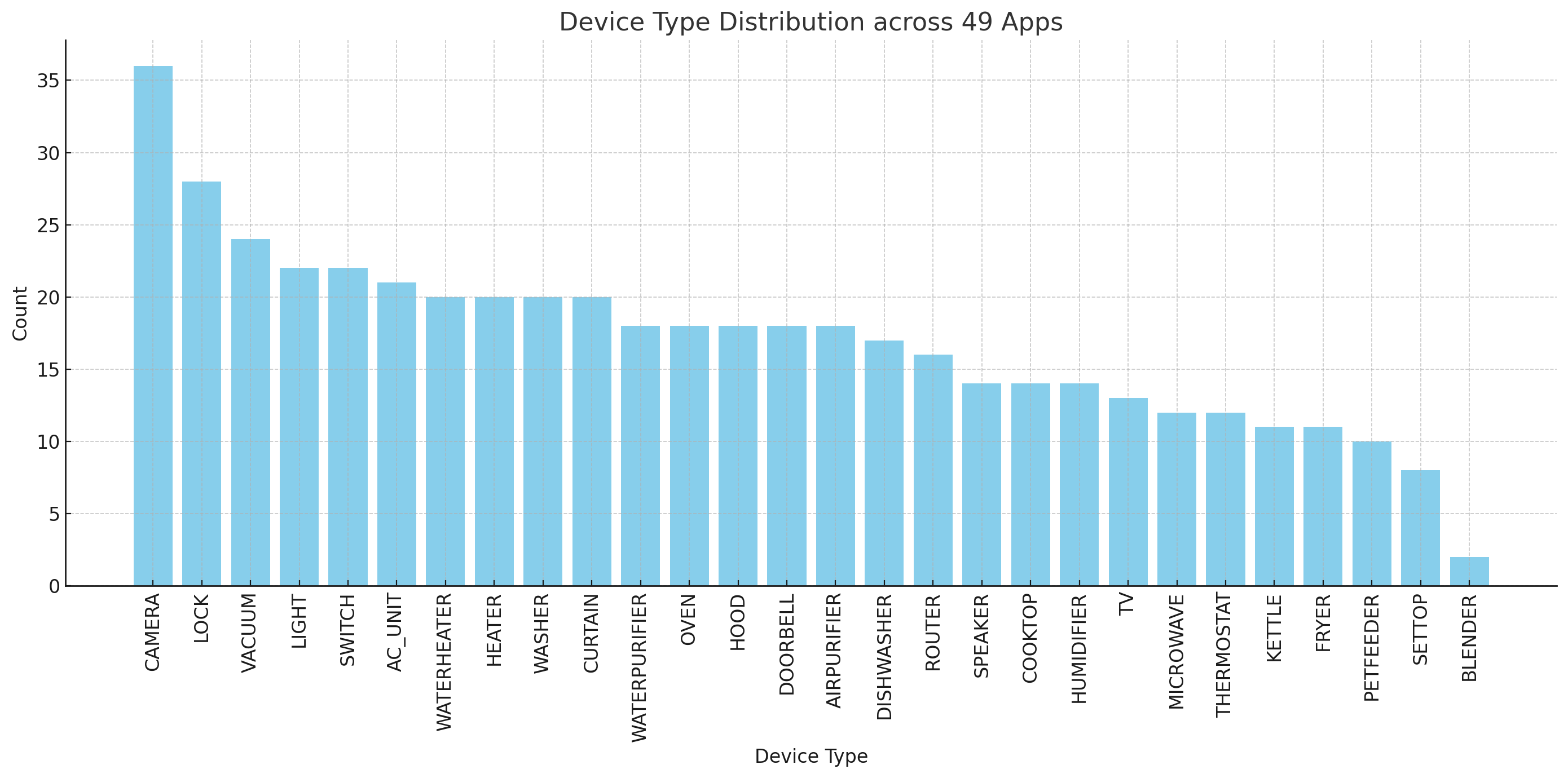}
\caption{Distribution of connected device types supported by Chinese smart home apps ($n=49$) as of January 2025.}
\label{fig:app_distribution}
\end{figure*}

Finally, we reviewed the 49 eligible apps (labeled APP1–APP49) and their developers to ensure that all were created by smart home companies.\footnote{We reviewed all these companies' sizes using the \href{https://bt.gsxt.gov.cn/index.html}{National Enterprise Credit Information Publicity System}.} The complete list of the 49 apps included in our analysis is available in \href{https://osf.io/v9e3j}{\textbf{this OSF repository}}. More than half of the apps ($n=25$) had been updated in January 2025, but only eight in-app privacy policies had been updated within the three months prior. Furthermore, based on Google's \href{https://developers.home.google.com/cloud-to-cloud/guides}{Smart Home Device Types} documentation, we categorized the apps in our study as follows: 36 apps could operate cameras and 27 could control door locks, both of which tend to pose higher bystander privacy risks~\cite{saqib2025bystander}, while only two apps could control blenders (see Fig.~\ref{fig:app_distribution}).
%\begin{table*}[ht]
    %\centering
    %\fontsize{7pt}{7pt}\selectfont
    %\begin{tabular}{lllll}
    %\toprule
    %\textbf{Company Type} & \textbf{Company Size} & \textbf{App Updated Time} & \textbf{Policy Updated Time} & \textbf{App Age Requirement} \\
    %\midrule
    %14\% SOE & 19\% 1-99 & 2\% March 2024 & 2\% Undisclosed & 77.6\% for 4+ 
    %\\
    %86\% PE & 6\% 100-499 & 2\% May 2024 & 2\% Jan-Dec 2020 & 4.1\% for 12+ 
    %\\
    % & 6\% 500-999 & 2\% June 2024 & 4.1\% Jan-Dec 2021 & 18.4\% for 17+  
    %\\
    % & 16\% 1000-4999 & 2\% Aug 2024 & 14.3\% Jan-Dec 2022 
    %\\
    % & 6\% 5000-9999 & 2\% Sep 2024 & 22.4\% Jan-Dec 2023 
    %\\
    % & 47\% 10000+ & 4\% Oct 2024 & 38.8\% Jan-Sep 2024 
    %\\
    % & & 6\% Nov 2024 & 16\% Oct-Dec 2024 
    %\\
    % & & 29\% Dec 2024 & 
    %\\
    % & & 51\% Jan 2025 & 
    %\\
    %\bottomrule
    %\end{tabular}
    %\caption{Company type and size, app and privacy policy update times, and age requirements for our selected smart home companies (by January 2025).}
    %\label{tab:app_and_company_info}
%\end{table*}}}}}}}%%%

%%

\subsection{Privacy Policy Analysis} \label{method_privacy_analysis}
We employed thematic analysis~\cite{braun2006using} to identify and report patterns in the collected privacy policy data, as this approach enables the systematic identification of themes related to privacy, security, and user rights. Data collection and analysis were conducted between January and June 2025. During the analysis and development of the codebook (described below), we refined our coding scheme through multiple iterative rounds to address the technical complexity of privacy policy language and the well-documented challenges of accurately coding such texts~\cite{mhaidli2023researchers}. Given the legal nature of privacy policies, we followed prior studies~\cite{javed2021privacy,malki2024exploring} by coding policies with reference to established regional privacy regulations and guidelines (e.g., GDPR). In this study, we used PIPL as the guiding framework, as it is the primary regulation governing the evaluation of mobile app privacy practices in China~\cite{calzada2022citizens}. Although PIPL does not define a distinct category for bystanders, its general principles and lawful bases still apply when apps and devices collect identifiable data about non-account holders in shared domestic spaces. In practice, this implies that privacy compliance in smart home ecosystems should extend beyond registered users to include anyone whose voice, image, or daily routines may be captured through cameras, microphones, or family-sharing features.%

We acknowledge that our dataset includes a wide range of smart home devices with varying data practices and bystander risk profiles, ranging from camera feeds and health data to simple temperature or lighting controls. However, our goal was to assess baseline compliance and consideration of bystander privacy across the Chinese smart home app ecosystem as a whole, using a general, principles-based legal framework (PIPL) that applies to all personal data controllers \textbf{regardless} of device type. Importantly, we found that explicit recognition of bystanders as data subjects in their own right, as well as corresponding bystander-facing privacy controls, was absent across all device categories (see \S\ref{result_data_control_management}). Thus, while we do not claim that all device types pose equivalent privacy risks to bystanders, our findings indicate that bystander privacy is consistently overlooked across app types, rather than being an artifact of any single class of smart home devices.

Following the PIPL framework, Authors 1 and 2 began by developing an initial codebook deductively. PIPL’s requirement that data controllers inform users about data collection methods directly informed our “Data Collection Methods” theme, which we further subdivided into specific codes, including “Direct Collection – Registration \& Real-Name” and “Indirect Collection – Device Identifiers.” The first author then applied the initial codebook to a randomly selected set of five privacy policies ($n=5$) to evaluate its effectiveness and revise it as needed. Subsequently, four authors (Authors 1–4, all native Mandarin speakers) independently coded another randomly selected set of five privacy policies (distinct from the initial set) using the revised codebook. Authors 1–4 then met to discuss their coding, resolve discrepancies, and merge their versions into a single consolidated codebook. Additional refinements were made to clarify category definitions, remove redundancies, and ensure alignment with both the codebook and the data. This iterative process was repeated two additional times until code saturation was reached, and the authors were confident that the codebook was comprehensive and well grounded in the data.%%%

After finalizing the codebook (see Table \ref{tab:codebook-en}), the first author coded all 49 privacy policies and discussed the coding with the other three authors, each of whom independently reviewed a subset of seven policies for consistency. Once coding was complete, all four authors collaborated to organize the codes into themes aligned with our research questions and reviewed the excerpts associated with each theme to further refine them.

We did not calculate inter-rater reliability (IRR) because our analysis followed an interpretivist, consensus-based approach in which the codebook was iteratively developed and refined. Qualitative outcomes can legitimately vary depending on researcher experience and the nature of the data; as Ortloff et al. noted, even when working with the same materials, different analysts may reach divergent yet equally valid interpretations, and IRR should not be treated as a proxy for quality in reflexive thematic analysis~\cite{ortloff2023different}. Following McDonald et al.'s recommendations~\cite{mcdonald2019reliability}, we established credibility through team reconciliation meetings rather than relying on a numerical agreement coefficient.%

\subsection{UX/UI Analysis} \label{method_ux_analysis}
We collected detailed UI screenshots and screen recordings from all 49 smart home apps, focusing on privacy-related features for both users and bystanders. These features included privacy settings, consent mechanisms, data collection notifications, and account management tools. The first two authors conducted evaluations independently between March and June 2025 using two iPhone 13 Pro devices, each with 256GB of storage and running iOS 18.2.1. No apps or iOS versions were updated during the testing period. Each app was systematically explored so that the researchers could be familiarized with its interface, features, and navigation. Task flows, interface elements, and any privacy-related features or notifications (e.g., for bystanders) encountered during both initial setup and regular use were documented.

\subsubsection{Cognitive Walkthrough} We employed a usability inspection method called cognitive walkthrough~\cite{blackmon2002cognitive} to assess interface usability and UX by simulating a first-time user's problem-solving process. Our analysis focused on tasks related to configuring privacy settings that affected bystanders, such as guests or other household members who are not the primary users. The primary user was defined as a novice, privacy-conscious individual using a Chinese smart home app, potentially lacking advanced technical skills. Their main goal was to understand and configure privacy and security settings for bystanders to prevent devices from infringing on others' privacy.

We designed five bystander-specific tasks and four general tasks (see Table~\ref{tab:privacy_tasks}). The first two authors independently simulated the user journey for each app, documenting detailed observations at every step---including design and usability issues, bystander configuration problems (e.g., lack of navigation feedback), unclear terminology, and barriers such as accidental shutdowns---using screenshots and screen recordings when necessary. Notes were then independently reviewed and compared in group discussions, applying constant comparison to identify recurring patterns and common issues.

Results were aggregated to uncover recurring patterns and themes, categorized by the first author. Each design issue was assessed from two perspectives: 1) its impact on the user's ability to manage their own privacy, and 2) its impact on the effective management of bystander privacy.

\subsubsection{Heuristic Evaluation and Privacy-by-Design Principles}
The same two authors independently conducted heuristic evaluations and applied privacy-by-design principles to assess how well each app supported bystander privacy, focusing on usability issues and privacy-related features. Nielsen's 10 Usability Heuristics~\cite{nielsen1990heuristic} served as the foundation (see Table~\ref{tab:heuristics}), supplemented by privacy-specific heuristics derived from privacy-by-design principles~\cite{cavoukian2009privacy} (see Table~\ref{tab:pbdd}).

Each author systematically navigated the apps, examining screens, menus, dialogs, and interactions related to privacy and security---for example, evaluating whether data practices were clearly communicated. Each issue was independently documented, noting the violated heuristic(s), a brief description, illustrative screenshots, and a severity rating (cosmetic, minor, major, or critical). This process also included any deceptive patterns identified~\cite{gray2018dark,deceptive_patterns}. After individual assessments, the authors reconciled discrepancies, consolidated similar issues, and categorized recurring problems, highlighting common usability barriers to effective privacy management.%

\subsection{Traceability Analysis} \label{method_traceablity}
To examine the alignment between declared policies and privacy-focused UI features, we conducted a manual traceability analysis~\cite{anthonysamy2013social,edu2021skillvet,misra2017privacy,edu2022measuring} across four key dimensions---\textbf{Data Collection}, \textbf{Data Sharing}, \textbf{Data Control and Management}, and \textbf{Security Measures}---from June to August 2025. Traceability analysis provides a systematic method for mapping stated policy provisions to implemented design features, revealing gaps that may undermine compliance, usability, or user trust.

We first extracted specific policy commitments related to these categories, including direct and indirect data collection (e.g., registration data, device identifiers, multimedia input), sharing practices (e.g., with third parties or law enforcement), user-facing control mechanisms (e.g., permissions, deletion options, bystander management settings), and technical safeguards (e.g., encryption). We then systematically evaluated how each app's UX/UI mapped to its corresponding policy commitments, based on observable elements in the app interfaces. The first two authors independently assessed each app and identified discrepancies between stated policies and implemented features, resolving differences through discussion until consensus was reached. Based on these results, we constructed a traceability matrix (see Table~\ref{tab:traceability_table}) linking privacy policies (rows) to observed UX/UI features (columns), coding each pair as \complete~\textit{Complete} (fully consistent and unambiguous matches), \notcomplete~\textit{Partial} (imprecise or ambiguous correspondences), \broken~\textit{Broken} (no matching UI controls for stated policy provisions), or \notapplicable~\textit{Not Applicable} (Neither policy nor UI elements were present).

We also collected and analyzed the Apple App Store privacy labels for each app (see Table~\ref{tab:privacy_labels}), motivated by prior work on standardized privacy notices and their usability~\cite{zhang2022usable,kelley2009nutrition}. These labels---self-reported by developers---summarize the types of data collected (e.g., data used to track users, data linked to users, data not linked to users) and the purposes for which they are used.\footnote{\url{https://developer.apple.com/app-store/app-privacy-details/}} We compared each app's Apple App Store privacy label with the app's privacy policy as well as its UX/UI features to identify inconsistencies, omissions, or potential misreporting in \S\ref{traceability_analyse}.%

\subsection{Author Positionality} \label{positionality}
Our positionality and interpretation of the results are shaped by our diverse backgrounds and experiences. All authors are experienced researchers in human-centered computing, privacy, and cybersecurity, with strong expertise in qualitative thematic analysis and extensive experience in UX evaluation. Author 4, although not specializing in HCI, contributed expert insights on legal frameworks and Chinese security policy.

While all authors are trained researchers currently working in predominantly Western institutions, the primary data analysts (Authors 1 and 2) brought additional cultural insights from their personal backgrounds, having been born and raised in China and being native Mandarin speakers. This sociocultural and legal familiarity enhanced their ability to interpret the data within the specific context of Chinese smart home environments. Such contextual understanding was particularly valuable for examining how privacy is constructed and negotiated in policy texts and interface designs, enabling the team to uncover underlying social norms, values, and ideologies---such as expectations around authority, collectivist regulations, family roles, and collective responsibility---that might not be apparent from a Western-centric perspective.

To minimize translation gaps between Mandarin and English, Author 1 first translated all privacy policies into English. Authors 2–4 then independently reviewed and manually revised the translations to ensure accuracy and clarity. The entire team subsequently discussed and confirmed that key terms---particularly technical and legal language---had been appropriately translated~\cite{xian2008lost}. We thus paid close attention to narrative structures and the interplay between visual and textual elements, enabling a deeper, contextually grounded interpretation.%%%%
\section{Results} \label{result}
\subsection{Privacy Policy Analysis} \label{privacy_policy}

\subsubsection{Privacy Policy Overview}
All 49 apps provided in-app privacy policies. Three companies released two separate apps each, with each app governed by a distinct policy. All apps stated that the policy scope covered the company’s hardware and software products, including smart home devices, companion applications, and associated online services. Each policy specified both update and effective dates (except APP25), with most updates taking effect on the same day they were issued. A total of 47/49 apps committed to notifying users of significant policy changes through pop-ups, emails, or public announcements; however, only six apps (e.g., APP9) explicitly described the types of changes that would trigger such notifications (e.g., the introduction of new features or revisions to data processing purposes). All apps reserved the right to update their policies, and most indicated that continued use of the service constituted acceptance of the revised terms.

Most apps (41/49) did not provide access to historical versions of their privacy policies. In contrast, some large companies (e.g., the developers of APP2, APP3, APP8, and APP9) not only documented policy version numbers but also maintained dedicated privacy websites that allowed users to review past changes and submit privacy-related concerns. Overall, 34/49 apps had updated their privacy policies at least once within the past 12 months, with APP4 and APP8 exhibiting the highest update frequencies. Apps that managed multiple devices—particularly camera-based systems or comprehensive smart home ecosystems—tended to update their policies more frequently. Conversely, some apps did not specify update intervals. For example, APP40 and APP13 were last updated in 2020 and 2022, respectively, potentially reflecting the narrower scope of their products (e.g., smart dishwashers) or a lower prioritization of privacy policy maintenance.

The average privacy policy length was 15,623 characters (SD = 6,575), with substantial variation across apps. Based on Wang et al.~\cite{wang2019visual}, who report a maximum reading speed of 259.5 $\pm$ 38.2 Chinese characters per minute, an average adult user would require approximately 52.5 to 70.6 minutes to read a typical policy in full. The shortest policy (APP46) contained 3,675 characters, whereas the longest (APP11) reached 34,226 characters. Most apps (40/49) exceeded 10,000 characters, employed formal legal language, and included technical terminology, posing notable comprehension challenges for non-expert users. All policies adopted hierarchical structures (e.g., chapters and sub-clauses) to facilitate navigation, and 38 apps used bold or colored text to emphasize key elements such as data categories or deletion rules for minors. Additionally, 11 apps enhanced clarity by linking to supplementary documents, including biometric data inventories (e.g., APP4) and third-party SDK lists (e.g., APP15). More than half of the apps (26/49) incorporated visual aids such as tables to illustrate permission scenarios (e.g., APP18, APP39), and 8 apps provided technical glossaries to support user understanding (e.g., APP5).%

\subsubsection{Data Collection Practices}
We categorized app data collection practices into two primary categories, guided by the requirements of relevant privacy laws and regulations (e.g., PIPL).

\textbf{Direct collection} primarily refers to information actively provided by users, including Personally Identifiable Information (PII) and registration details that require manual input by the primary user. All apps required primary users to submit basic information to register accounts and activate devices, most notably phone numbers.\footnote{According to Art. 24 of the CSL, China’s real-name system requires individuals to use their actual identities to obtain phone numbers and Internet services and to access online platforms. If users do not provide PII, network operators are prohibited from offering the corresponding services.} Optional information included usernames, avatars, and passwords, which also relied on direct user input. However, most apps broadly stated that they collected data within the scope authorized by users, without specifying the exact data types involved.

\textbf{Indirect collection} involves data gathered through device resources and user interactions, the integration of third-party apps or SDKs, and tracking technologies such as cookies. All apps consistently collected device-specific identifiers, including IMEI and MAC addresses, operating system versions, and network information (e.g., Wi-Fi SSIDs). They also collected precise or approximate location data (e.g., latitude and longitude coordinates, IP addresses, and base station information) and passively gathered data via cookies. All 49 apps required network-related data to establish device connectivity and requested multimedia permissions (e.g., camera or microphone access) to enable device activation. Beyond basic functionality, apps also collected logs and interaction data to enhance UX. Some apps further leveraged automation logs, energy consumption data, or device-specific operational metrics to support smart home–specific scenarios. For instance, APP26 analyzed more than six months of historical data, including temperature, humidity, and energy usage, to generate daily life reports.

\textbf{Sensitive information.} All apps explicitly described the scope and use scenarios associated with the collection of sensitive personal information. While baseline functionalities and compliance requirements were largely standardized, implementation strategies, particularly for handling sensitive biometric data and deploying AI-driven personalization features, varied substantially across apps. As part of core functionality, 39/49 apps reported collecting users’ financial information, such as payment amounts, bank card details, or third-party payment account information, which were collected only during transaction processes, as well as search histories and shopping records. Regarding biometric data, all apps stated that they collected biometric information, including facial recognition data (e.g., via camera-based devices). Some apps additionally reported collecting health-related metrics, such as blood pressure, blood sugar levels, menstrual cycle data, movement activity, heart rate, weight, and sleep patterns. Other forms of biometric data included fingerprints, vein patterns, and voiceprints. Notably, although these apps might directly or indirectly collect data from bystanders (e.g., capturing the face of a visiting friend or passerby via a smart doorbell), none of the apps explicitly disclosed the collection mechanisms used or addressed how sensitive personal information related to bystanders was handled.%%%

\subsubsection{Data Storage and Sharing Practices} \label{subsub_data_sharing}
We highlight all apps' strict compliance with China’s data localization requirements and identify three distinct storage models based on their approaches to cross-border data transfers under PIPL. We also describe data sharing practices involving live-in bystanders (e.g., family members), third-party service providers (e.g., advertisers), and law enforcement agencies (e.g., public security authorities).

\textbf{Data transfer and sharing methods.} All apps demonstrated strict adherence to China’s data localization requirements and referenced the principle of retaining or storing only the minimum necessary data. Additionally, they clearly distinguished between domestic third-party data sharing (e.g., with logistics providers, payment processors, or cloud vendors) and cross-border transfers, the latter requiring explicit user consent as mandated by PIPL. Across the apps, we identified three distinct data storage models.

The first model is a strict onshore-only model, retaining all user data within mainland China without routine cross-border transfers. APP28 stated: \textit{``Personal information collected within China will be stored domestically and will not be transferred outside of China unless required by law.''}

The second model involves onshore-first storage with conditional outbound transfers, allowing data to be sent abroad only under exceptional circumstances, subject to user consent, encrypted transmission, and a security assessment by the Cyberspace Administration of China (CAC). APP26 noted: \textit{``Personal information collected and generated within mainland China will be stored domestically. Where business or legal requirements necessitate cross-border transfer, we will comply with applicable laws and regulations, enter into the required agreements, conduct security assessments or audits as necessary, and require overseas recipients to implement appropriate measures to ensure the information confidentiality.''}

The third model features geo-distributed or globally deployed storage, supporting continuous data exchange across international cloud infrastructures. This model was adopted by companies with overseas operations (e.g., developers of APP12, APP21, APP47), which operated multiple regional cloud servers and cited compliance with frameworks such as the EU’s GDPR. For instance, APP22 referenced GDPR mechanisms and pledged equivalent protections, alongside mandatory filings with CAC. Some apps, such as APP39, adopted a relay-only passthrough model, in which real-time media was temporarily transmitted via overseas servers without permanent storage, stating: \textit{``Real-time audio and video data are transmitted overseas solely for temporary relay, without any permanent storage, and therefore do not constitute a cross-border transfer of personal information under PIPL.''}%%

\textbf{Data sharing with third parties.} All policies mentioned sharing user PII with third-party business partners and service providers, for purposes such as R\&D improvements and advertising. For example, APP16 stated: \textit{``We will use personally identifiable information (PII) collected in accordance with applicable laws to generate algorithmic models that predict your preference profile. Based on these preferences, we may recommend products and services that are likely to interest you, while also offering options that are not personalized.''} However, only 12/49 policies clarified how users could manage such sharing, for instance by adjusting or disabling PII collection or opting out of third-party advertising. This lack of transparency and control raises concerns about user agency and informed consent, particularly in the context of algorithmic profiling. To mitigate such concerns, some apps emphasized privacy-preserving data-sharing strategies (e.g., aggregated or anonymized data). Nonetheless, the extent to which anonymization was applied, and whether it met technical or regulatory standards, remained unclear in most policies.

All policies explicitly acknowledged that data sharing with public security authorities (e.g., police, national security departments) was mandatory when required. Article 28 of CSL mandates that network operators and service providers must provide technical support and assistance to law enforcement and national security agencies in accordance with the law. Consequently, smart home companies are obliged to retain and disclose data related to criminal investigations, national security concerns (e.g., public safety or health), or illegal content (e.g., political dissent, violence, or terrorism). This prioritization of state interests over individual rights was evident across all policies. While these provisions are framed as protecting public welfare, they effectively deprioritize user privacy when national security or public interest is involved. In particular, the broad and vague definitions of \textit{``public interest''} or \textit{``national security''} grant authorities wide interpretive discretion, creating a legal environment where personal data, especially from smart home surveillance, can be accessed without user awareness or consent. For instance, APP10 stated: \textit{``When necessary to fulfill statutory obligations; when necessary to respond to public health emergencies or to protect the life, health, or property of natural persons in emergencies; when personal information is processed within a reasonable scope for news reporting, public opinion supervision, or other public-interest activities; or in other circumstances provided by laws and administrative regulations.''}

\textbf{Data sharing with other users (live-in bystanders).} 19/49 policies provided explicit mechanisms for users to share data with other users, typically other household members. Although this group could be considered live-in bystanders~\cite{saqib2025bystander}, the policies did not identify or define any other users or household members as ``bystanders.'' While the policies often emphasized configurability and voluntary participation, they generally lacked specific provisions to safeguard the privacy of live-in bystanders. For instance, APP4 and APP10 allowed users to enable voiceprint recognition that captured audio data from both the primary user and other household members to deliver tailored services. Some apps enabled users to invite household members through third-party platforms such as WeChat to facilitate device sharing. APP10 supported family health functions that collected sensitive biometric and medical information (e.g., heart rate, sleep patterns) to generate wellness recommendations, allowing users to submit personal data on behalf of household members with their consent. A few apps (e.g., APP15) collected contact and identity information of household members when users enabled features such as remote unlocking for smart locks. However, most apps did not clearly define how consent should be obtained from secondary users or how their data would be managed independently. In shared household settings, where roles, power dynamics, or technical literacy may vary, this absence of granular control and clarity introduces potential risks, such as unintentional data exposure or misuse by other household members.%%

\subsubsection{Data Control and Management Practices} \label{result_data_control_management}
Despite all policies recognizing users' data protection rights, many apps offered deletion and purging mechanisms that were often constrained by public security mandates. Data export and retention controls were fragmented, and protections for bystanders were inconsistently mentioned or entirely overlooked.

\textbf{Data and account deletion methods.} Most policies supported user-initiated data deletion alongside automatic data purging. However, users' deletion rights could be overridden by judicial investigations (as per Art. 28, CSL), rendering deletion requests effectively invalid. This may create legal conflicts with Art. 47, PIPL, which guarantees users the right to delete personal data. Yet, since CSL takes precedence, companies must comply with government directives. For example, APP17's township-version authorization included a public security mandate: video feeds from residents' cameras were stored indefinitely on servers operated jointly with local governments, and deletion was prohibited without official approval. This arrangement also granted government personnel full, real-time control over user devices, enabling unfettered access and significantly increasing the risk of privacy violations. Although PIPL (Art. 23) nominally requires separate, explicit consent for third-party access, enforcement remains largely symbolic; once users share control, township platforms may continue collecting and retaining footage even after users delete their accounts.

Data deletion workflows varied notably across apps. First, all apps offered account deletion to meet compliance requirements, and most provided straightforward, single-step self-service options. For example, APP31 provided a clear in-app pathway for account deletion, stating: \textit{``You can immediately delete all your information via the app by following this path: My Settings → Account and Security → Account Cancellation.''} Second, some apps implemented dual-level controls, enabling separate deletion paths for device-specific data and full account erasure. Third, system-driven expiry models were common among apps handling large volumes of media content, such as APP2, APP8, and APP27. These apps automatically purged videos and voice recordings after predefined periods (7–30 days), with logs rotating every six months and backups cleared during later cycles. Notably, only a few apps specified that if accounts were canceled, all personal data---especially children’s data---would be erased or anonymized (e.g., APP6).

\textbf{Data retention.} Only eight apps mentioned data retention practices such as data export and the right to data portability. These apps typically described two methods for data retrieval: direct export by users through in-app settings, or submitting a request via customer service. However, none of the apps supported full data downloads, and only APP44 specified the types of data available for export, stating: \textit{``Click My Settings → Account and Security → Export Account Information → Request Export, and enter the email address where you would like to receive your data. After verifying your account, we will send the following PII to the provided email address: your registered mobile number, username, profile photo, and the list of devices linked to your account.''}%%%%

\textbf{Age restrictions for children.} To comply with the \href{https://www.gov.cn/gongbao/content/2019/content_5456808.htm}{Provisions on Cyber Protection of Children's Personal Information (PCPCPI)}, most apps (47/49) specified what data of children (i.e., individuals under 14 years old) was collected. These apps explicitly collected data such as birth dates, gender identities, names, schools, and academic records. For example, APP6 emphasized that their policies did not apply to minors under a certain age, providing separate ``Guidelines on the Protection of Children’s Personal Data,'' and noting special responsibilities for children’s guardians in alignment with Art. 5 of PCPCPI: \textit{``As a guardian, you should properly fulfill your guardianship duties, educate and guide your child to raise their awareness and ability to protect personal data, and ensure the security of their personal data. You should also accompany your child in reading and agreeing to this guideline.''}

\textbf{Deceased users’ data management.} PIPL grants close relatives the right to manage the personal data of deceased individuals (Art. 49). However, most apps (45/49) lacked explicit provisions regarding the privacy of deceased users. For instance, APP47 stated: \textit{``We will protect the personal information of deceased individuals in accordance with the provisions of the PIPL. After a user’s death, their close relatives may, for their own lawful and legitimate interests, exercise rights such as access, copying, correction, and deletion of the deceased user’s personal information—unless the user made other arrangements during their lifetime.''}

\textbf{Bystander data management.} Many apps (19/49) introduced data sharing for other users (live-in bystanders) (see \S\ref{subsub_data_sharing}). However, policies generally did not provide clear pathways for exercising rights or explicitly reference legal standards regarding bystanders' privacy. The remaining 30 apps were vague or entirely omitted critical scenarios involving bystander protections. Only a few apps specified user types; for example, APP32 distinguished between end-users and bystanders: \textit{``Some models of home robots support a device sharing function, allowing the paired app account holder (the “sharer”) to share control rights with other app account holders (the “sharees”).''} APP2 further mentioned integration with the app account system, covering functions such as smart device linking and family sharing.

Although Art. 13 of PIPL mandates obtaining separate consent for processing another individual's data, only a few apps addressed user responsibility. APP24 warned: \textit{``Please exercise caution when uploading videos containing others’ sensitive information; otherwise, the uploader may bear legal responsibility.''} Some apps, like APP29, simply advised users to \textit{``avoid including other people’s private information.''} In some doorlock apps, such as APP31, data sharing required inputting family members' phone numbers but omitted specific bystander deletion rights. The lack of direct control for bystanders over their data poses significant privacy risks; for instance, visitors or neighbors inadvertently captured in surveillance footage have limited recourse for deleting their data.%%%

\subsubsection{Security Measures} All app policies outlined a combination of technical protections and organizational management practices to safeguard user data. Commonly cited measures included encryption, strict access control mechanisms, regular employee training, proactive incident response mechanisms, and systematic auditing processes. Most apps also included disclaimers about the practical limits of data security, emphasizing that responsibility for protection was shared between companies and users.

\textbf{Technical measures.} All apps emphasized safeguarding personal data through technical measures, typically aligning with industry-standard security practices. For example, APP15 noted: \textit{`We encrypt personally identifiable data during transmission and storage to ensure confidentiality. For mobile device identifiers such as IMEI and MAC addresses, anonymization and encryption using the MD5 algorithm are performed on users’ devices before collection and upload. Data that is displayed or correlated is further processed using multiple data-desensitization techniques, including SHA256.''} Similarly, APP47 stated: \textit{`We employ specialized access-control measures for cloud-based data storage [...]. Data exchanged between user devices and our servers is protected using SSL and other encryption algorithms [...]. We also regularly review our data handling practices to prevent unauthorized access.''}

Furthermore, 18/49 apps stated adherence to the \href{https://www.gov.cn/}{Management Rules for Multi-Level Protection of Information Security} and national standards like \href{https://openstd.samr.gov.cn/bzgk/gb/newGbInfo?hcno=BAFB47E8874764186BDB7865E8344DAF}{GB/T 22239-2019}, which apply when the destruction of an information system would cause serious harm to social order, public interests, or national security. Only 9/49 apps referenced established security management frameworks aligned with international standards such as \href{https://www.iso.org/standard/27001}{ISO 27001}.%%%

\textbf{Management measures.} Alongside technical safeguards, management practices formed a key component of data protection strategies. A common approach was minimizing access permissions, with most policies endorsing the \textit{``principle of minimal necessary access,''} ensuring that only authorized personnel handle sensitive data. For example, APP11 noted: \textit{``Trusted protection mechanisms against malicious attacks, along with strict access-control measures, ensure that only authorized personnel can access personal information.''}

Many apps also described comprehensive incident-response mechanisms, outlining contingency plans for data breaches and emphasizing training as a core component of management. For instance, APP26 highlighted the importance of \textit{``conducting regular security and privacy training to enhance employee awareness.''} Several apps adopted vetting procedures and contractual safeguards for third-party data handlers, with some explicitly stating that they conducted \textit{``due diligence to ensure that third-party service providers and business partners have robust measures in place to protect personal data,''} supported by regular audits.

Despite these assurances, all apps acknowledged the inherent risks of Internet-based data transmission and storage. Nearly every policy included liability disclaimers, noting that absolute security could not be guaranteed. Some apps explicitly warned that \textit{``the Internet cannot be guaranteed to be completely secure,''} urging users to take proactive steps, such as using strong passwords and updating them regularly, to complement organizational safeguards.%%

\begin{shaded}
\textbf{RQ1:} Chinese smart home apps frequently emphasized compliance with national laws and regulations (e.g., PIPL) and described privacy practices in their policies, including technical safeguards (e.g., encryption), management protocols (e.g., employee training), and distinctions between domestic and cross-border data sharing. Many apps framed security as the primary concern, sometimes prioritizing protection against external threats over individual privacy, which could inadvertently justify invasive practices. Despite the existence of governance mechanisms to protect primary users, bystander privacy remained insufficiently addressed. 19 out of 49 apps recognized scenarios that could adversely affect live-in bystander privacy in smart homes, yet most lacked explicit consent procedures or data deletion rights for bystanders. This gap highlights a broader pattern: although smart home apps implemented strong security and user control measures, they often overlooked the risks in shared domestic environments, where bystanders could be passively surveilled without consent or recourse.
\end{shaded}%%%

\subsection{UX/UI Analysis} \label{ux_analyse}
\subsubsection{Account Creation} \label{UX_account_creation}
\begin{figure*}[htbp]
    \centering
    \begin{subfigure}[t]{0.22\textwidth}
        \centering
        \includegraphics[width=\textwidth]{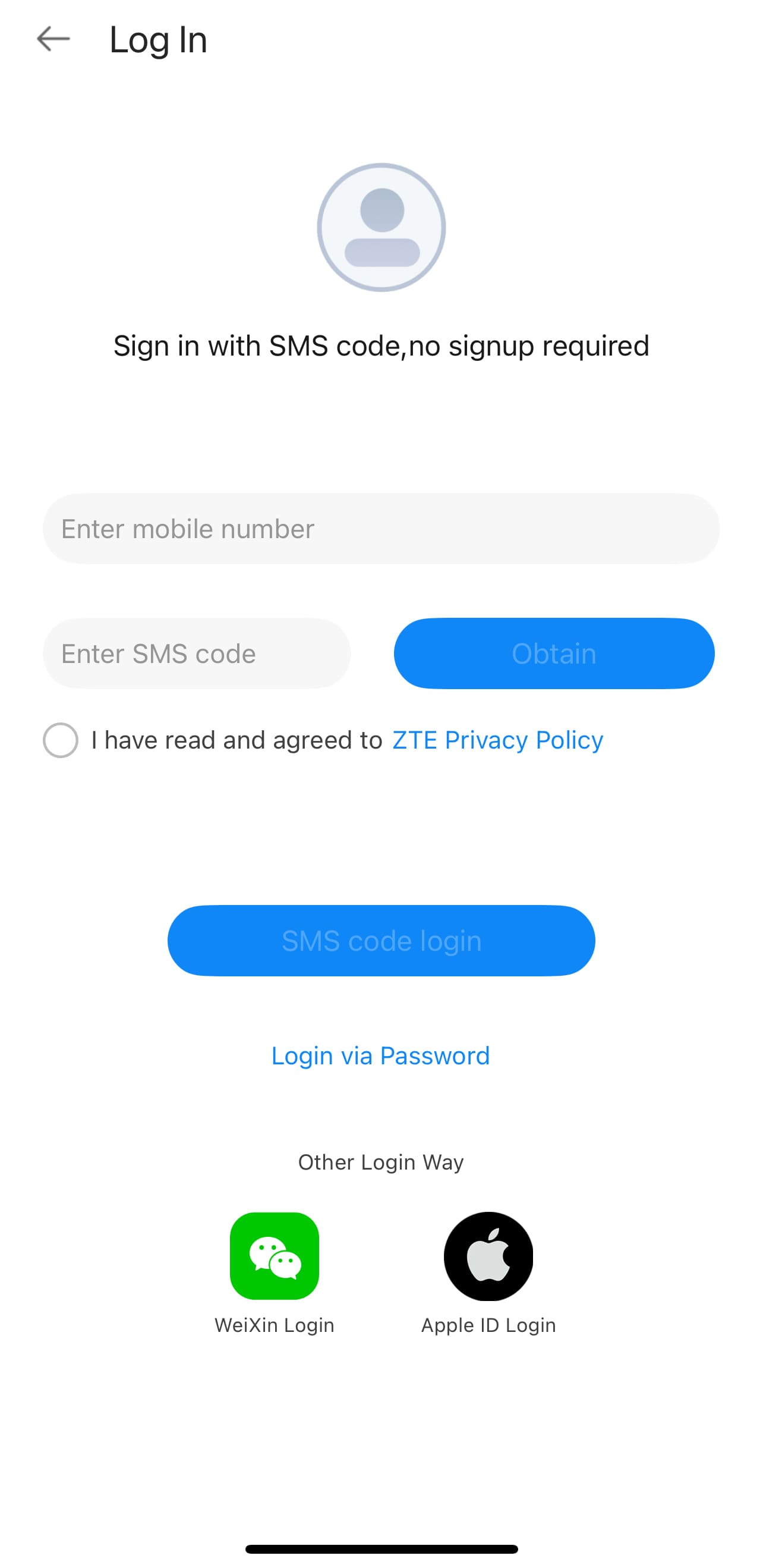}
        \caption{Real-name registration with SMS verification code (APP23).}
        \label{fig:1b}
    \end{subfigure}
    \hspace{0.02\textwidth}
    \begin{subfigure}[t]{0.22\textwidth}
        \centering
        \includegraphics[width=\textwidth]{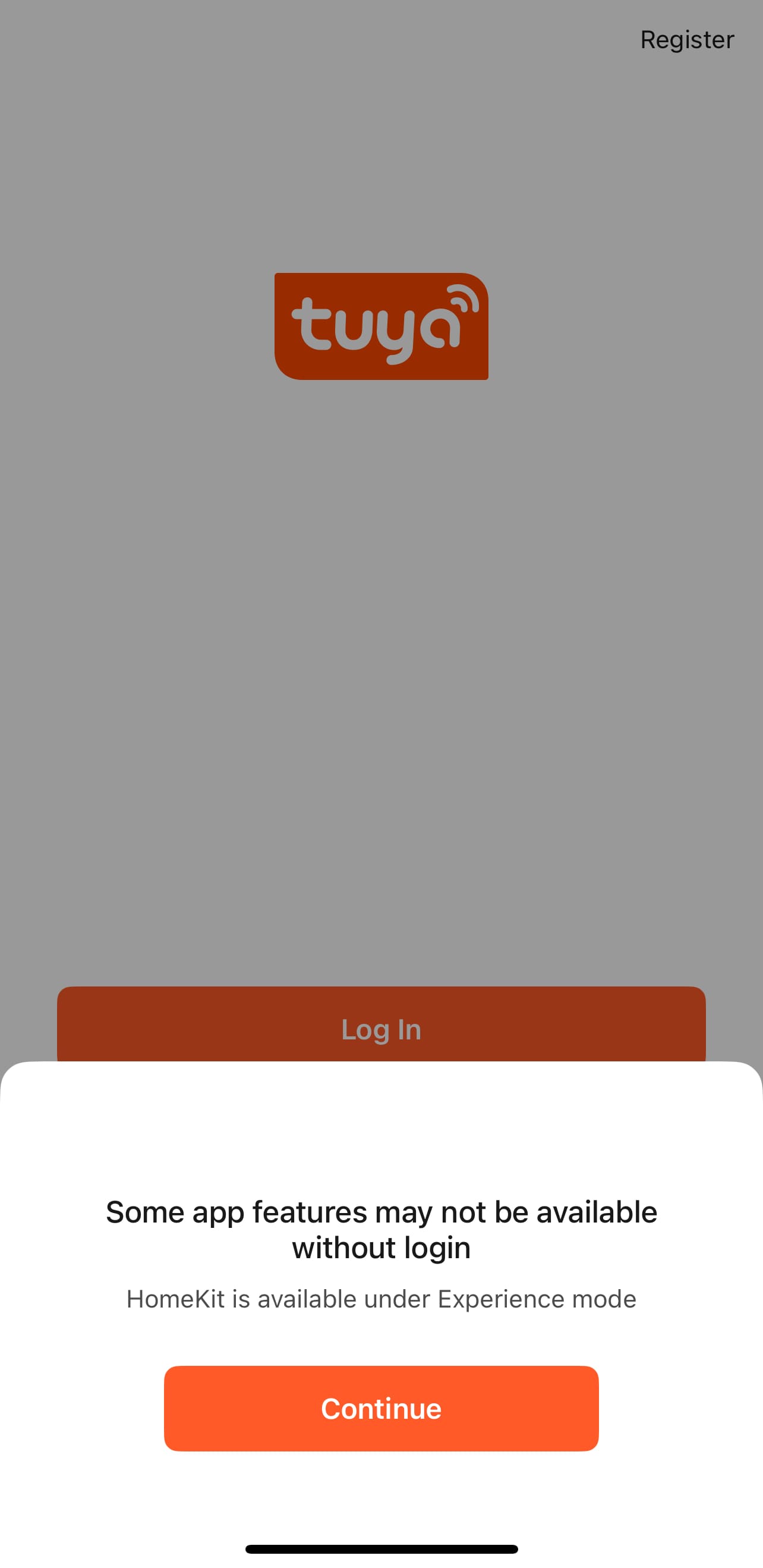}
        \caption{Onboarding page with guest mode option and limited features (APP21).}
        \label{fig:1c}
    \end{subfigure}
    \caption{Examples of app registration and onboarding screens.}
    \label{fig:account_creation}
\end{figure*}

All apps required users to register with a mobile number and verify their identity via a one-time code, in compliance with China’s real-name system (see Fig.~\ref{fig:1b}), and all prominently displayed a \textit{Terms of Use} notice with a consent mechanism---typically a checkbox---requiring acceptance of the full privacy policy before access. In the iOS environment, registration patterns varied: most apps (37/49) offered mainstream sign-up options such as WeChat or Apple ID, 11/49 restricted registration to mobile numbers only, some supported third-party social logins like Weibo (7/49), and three provided Google sign-in options. Two apps additionally offered a guest or experience mode that allowed limited access without account creation, with certain features restricted (see Fig.~\ref{fig:1c}). Despite these differences, users were generally not informed at registration about the specific types of data collected or which data was essential for basic functionality, and from a bystander perspective, no app disclosed how data related to non-account holders would be collected or managed, leaving bystanders entirely uninformed about potential data capture at the account creation stage.%%

%%%

\subsubsection{User Privacy Controls} \label{UX_privacy_controls}
\begin{figure*}[htbp]
    \centering
    \begin{subfigure}[t]{0.22\textwidth}
        \centering
        \includegraphics[width=\textwidth]{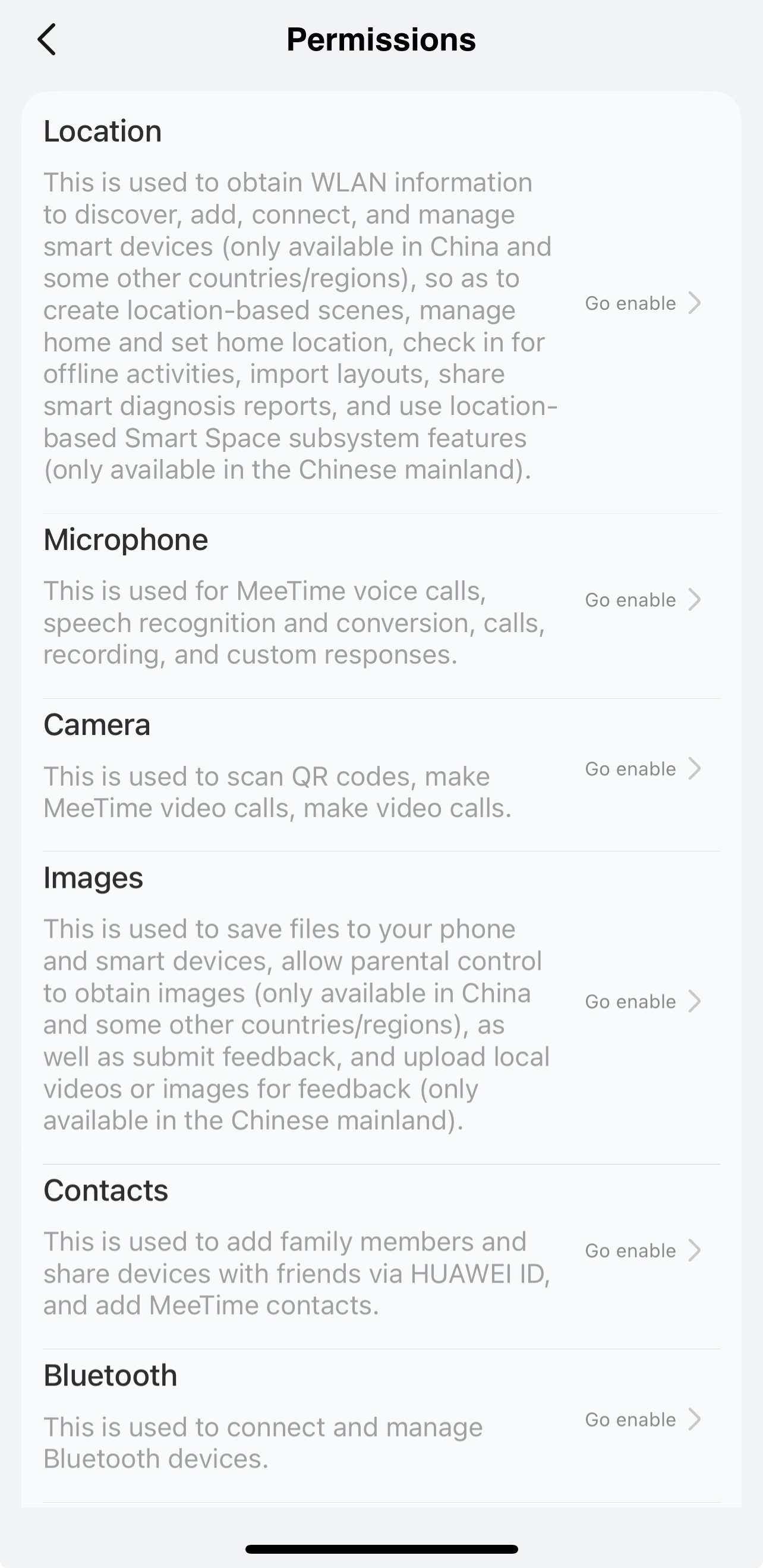}
        \caption{Detailed explanations of each permission (APP9).}
        \label{fig:2a}
    \end{subfigure}
    \hspace{0.02\textwidth}
    \begin{subfigure}[t]{0.22\textwidth}
        \centering
        \includegraphics[width=\textwidth]{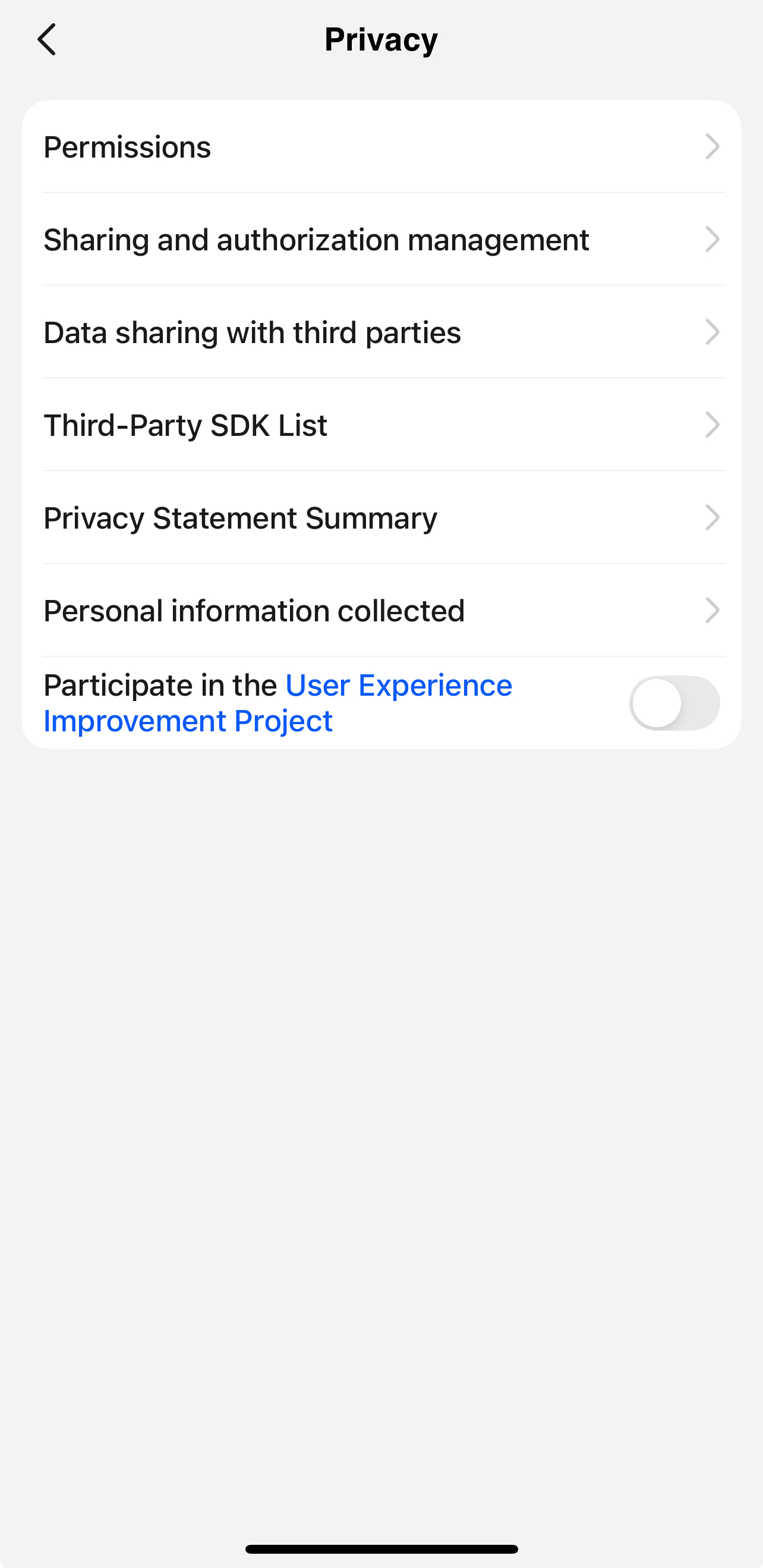}
        \caption{Settings accessible via direct in-app toggle controls (APP9).}
        \label{fig:2b}
    \end{subfigure}
    \hspace{0.02\textwidth}
    \begin{subfigure}[t]{0.22\textwidth}
        \centering
        \includegraphics[width=\textwidth]{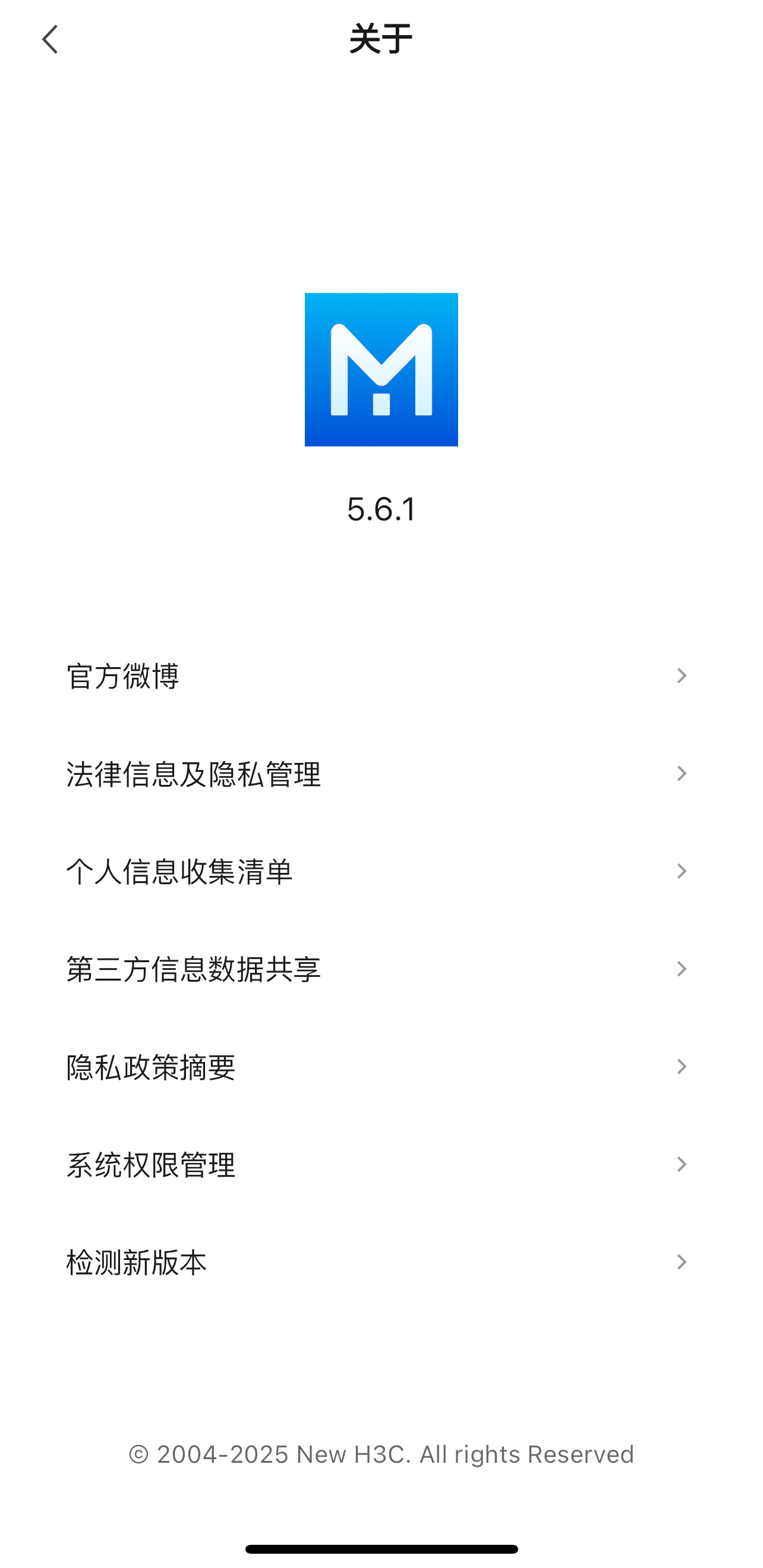}
        \caption{Privacy settings located in less accessible sections (APP36).}
        \label{fig:2c}
    \end{subfigure}
    \hspace{0.02\textwidth}
    \begin{subfigure}[t]{0.22\textwidth}
        \centering
        \includegraphics[width=\textwidth]{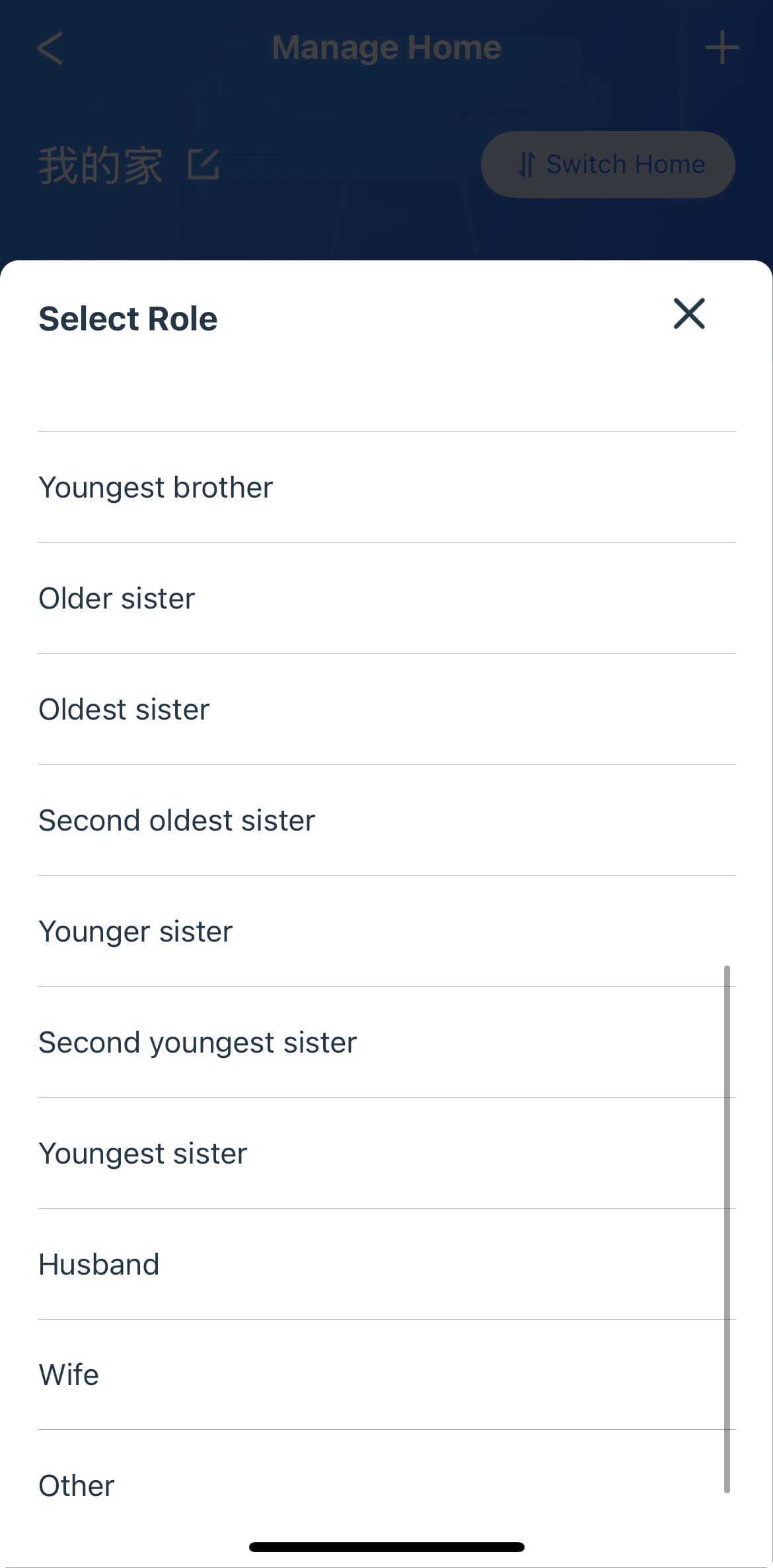}
        \caption{Explicit settings for managing device use and sharing with live-in bystanders (APP32).}
        \label{fig:2d}
    \end{subfigure}
    \caption{Examples of in-app privacy controls and mechanisms for managing bystander privacy.}\label{fig:privacy_controls}
\end{figure*}

Across all selected apps, \textit{user privacy control} emerged as a broad concept encompassing system permission management, data sharing preferences, and personalized recommendation settings---typically integrated within a unified settings interface. A total of 28/49 apps offered granular privacy controls, enabling users to customize settings with high usability. For example, APP9 provided clear, detailed explanations for each permission request (see Fig.~\ref{fig:2a}), and related options could be adjusted directly via in-app toggles (see Fig.~\ref{fig:2b}). However, five apps featured low-usability designs, where privacy settings were buried in less accessible sections such as \textit{About the App}, making navigation unintuitive (see Fig.~\ref{fig:2c}). Additionally, 16/49 apps lacked visible or clearly defined privacy control interfaces altogether, requiring users to manually adjust permissions via the iOS Settings app (e.g., APP39).

Regarding bystander privacy, only a few apps, such as APP32, provided explicit settings for managing device sharing with bystanders, clearly outlining the types of live-in bystander roles that could be assigned (see Fig.~\ref{fig:2d}). Conversely, most apps lacked dedicated bystander privacy controls, leaving data management entirely under the primary account holder's authority and denying bystanders any direct agency over their own privacy settings.%%%

%%

\subsubsection{App Permission Collection} \label{UX_permission_collection}
\begin{figure*}[htbp]
    \centering
    \begin{subfigure}[t]{0.22\textwidth}
        \centering
        \includegraphics[width=\textwidth]{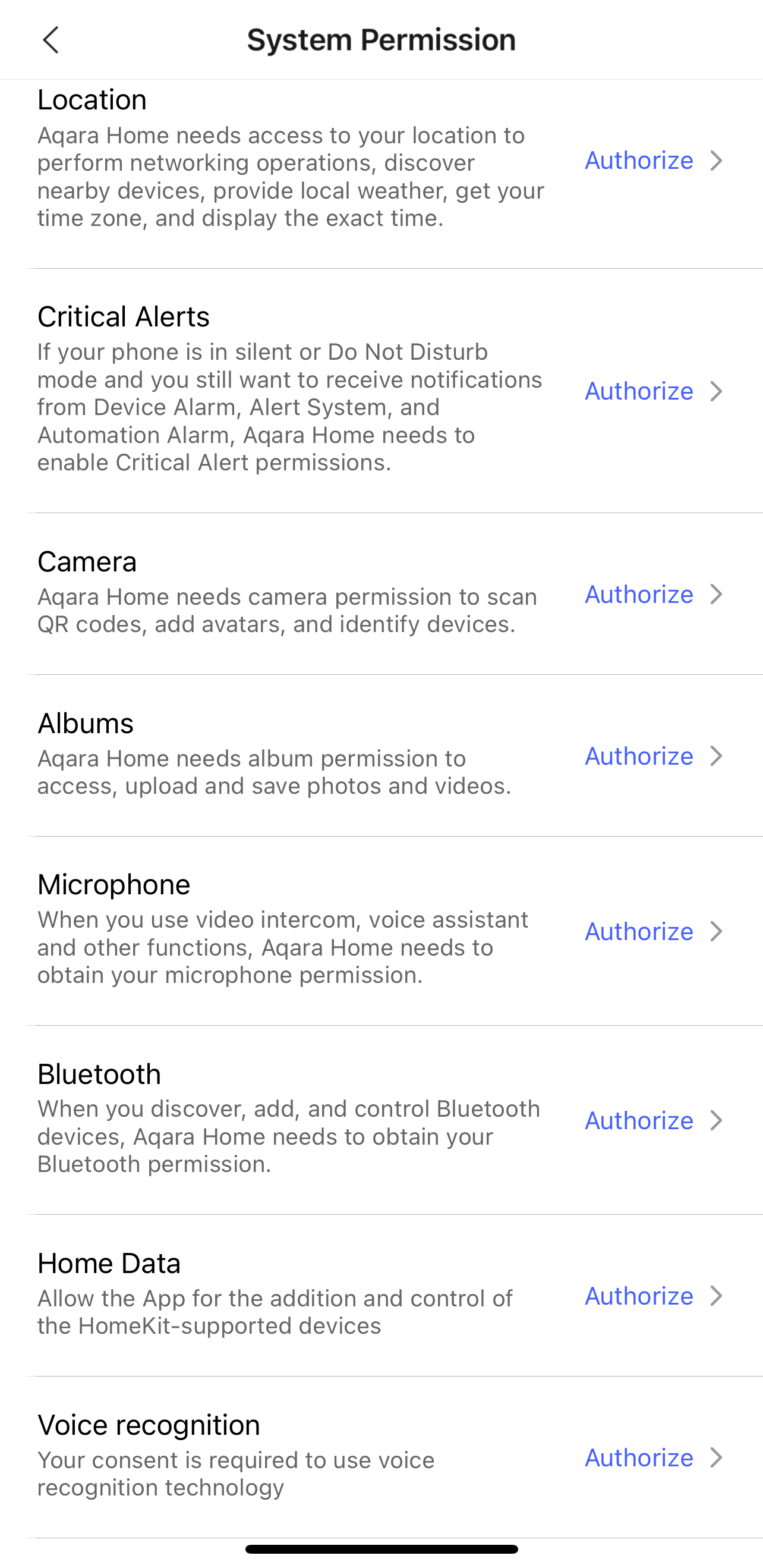}
        \caption{Core permissions required to enable full functionality (APP26).}
        \label{fig:3a}
    \end{subfigure}
    \hspace{0.02\textwidth}
    \begin{subfigure}[t]{0.22\textwidth}
        \centering
        \includegraphics[width=\textwidth]{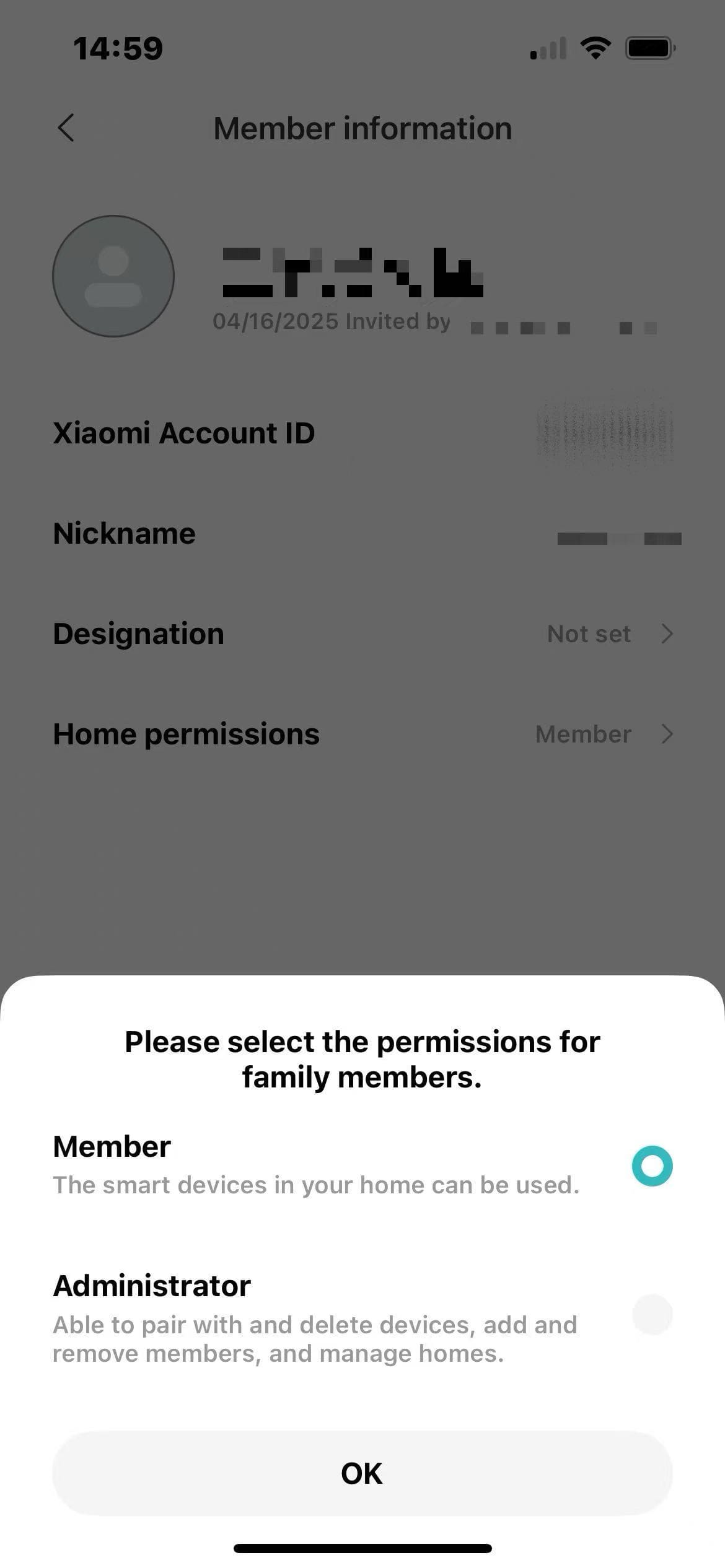}
        \caption{Primary users modifying bystander permissions without bystander consent (APP2).}
        \label{fig:3b}
    \end{subfigure}
    \caption{Examples of app system permissions and controls.}
    \label{fig:permission_collection}
\end{figure*}

Permission requests varied depending on the connected device type, resulting in noticeable differences across apps. Most apps collected extensive user data via seven core permissions---location, camera, photo album, microphone, contacts, Bluetooth, and system notifications---to ensure full service functionality (see Fig.~\ref{fig:3a}). Interfaces for permission requests typically included brief explanations of each request’s purpose: location enabled scene automation, home management, device discovery, and time synchronization; camera allowed QR code pairing; photo album access supported media uploads; microphone enabled voice assistant features; contacts allowed family management; Bluetooth facilitated device connectivity; and system notifications ensured timely alerts and updates. Notably, 25/49 apps integrated e-commerce features, which requested location access to enable delivery of home appliances, smart devices, and accessories. Some apps also requested less conventional permissions: APP4 and APP34 accessed calendars for reminders; phone access enabled customer support contact; APP15 used storage to save recorded media; APP16 tracked fitness data; APP17 used overlay permissions for live camera feeds; APP18 enabled Face ID for authentication; and APP26 accessed home data for broader HomeKit integration. From a bystander perspective, most apps did not explicitly request permissions for capturing or storing data about non-users inadvertently recorded by device functionalities (e.g., visitors in camera footage). Only a few apps (e.g., APP2) provided permission-related controls for bystanders; however, bystanders themselves could not directly configure or edit these controls, which were instead managed by the primary users (see Fig.~\ref{fig:3b}).%%%

\begin{figure*}[htbp]
    \centering
    \begin{subfigure}[t]{0.22\textwidth}
        \centering
        \includegraphics[width=\textwidth]{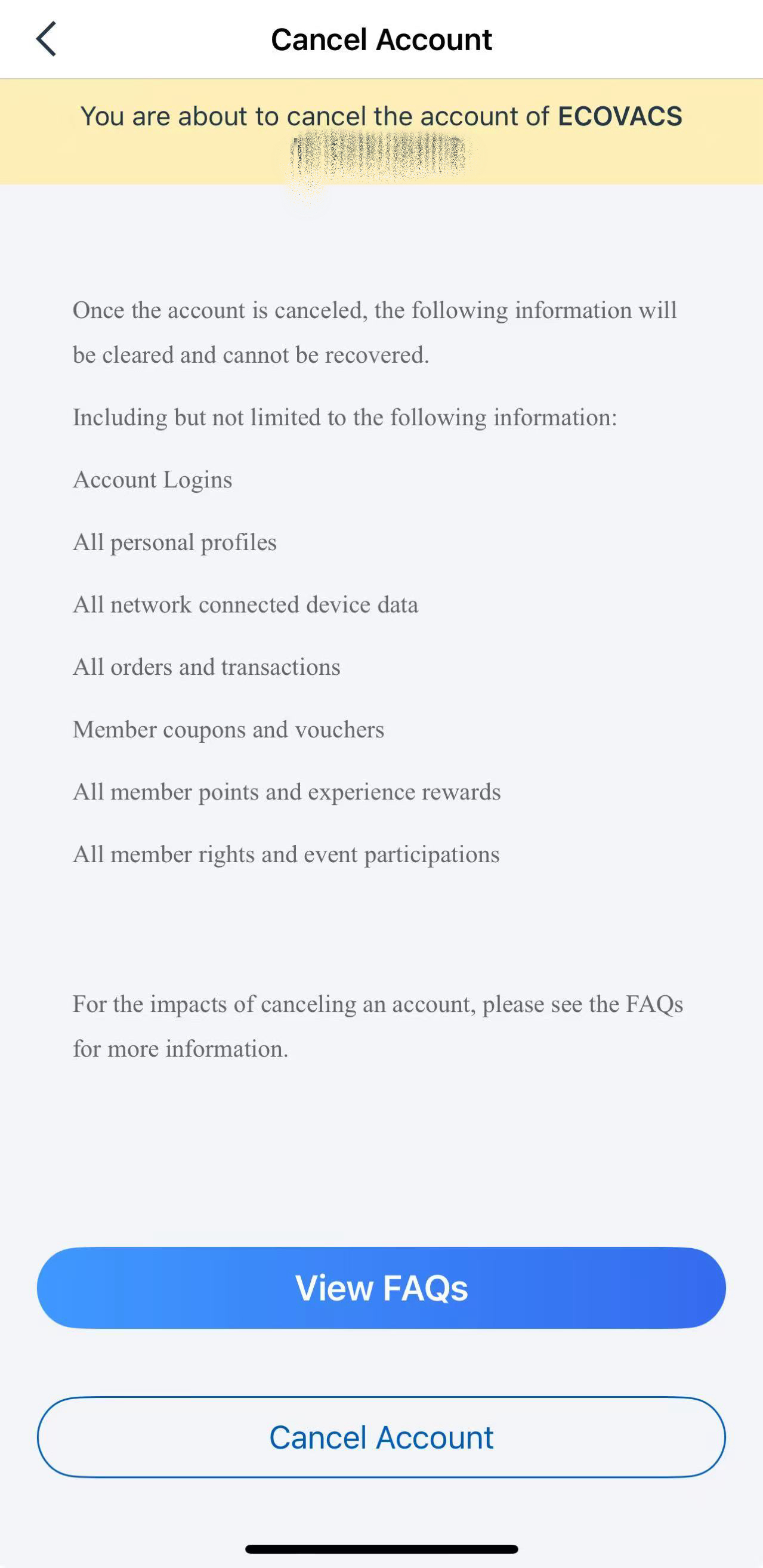}
        \caption{Single deletion option for complete, irreversible account removal (APP32).}
        \label{fig:4d}
    \end{subfigure}
    \hspace{0.02\textwidth}
    \begin{subfigure}[t]{0.22\textwidth}
        \centering
        \includegraphics[width=\textwidth]{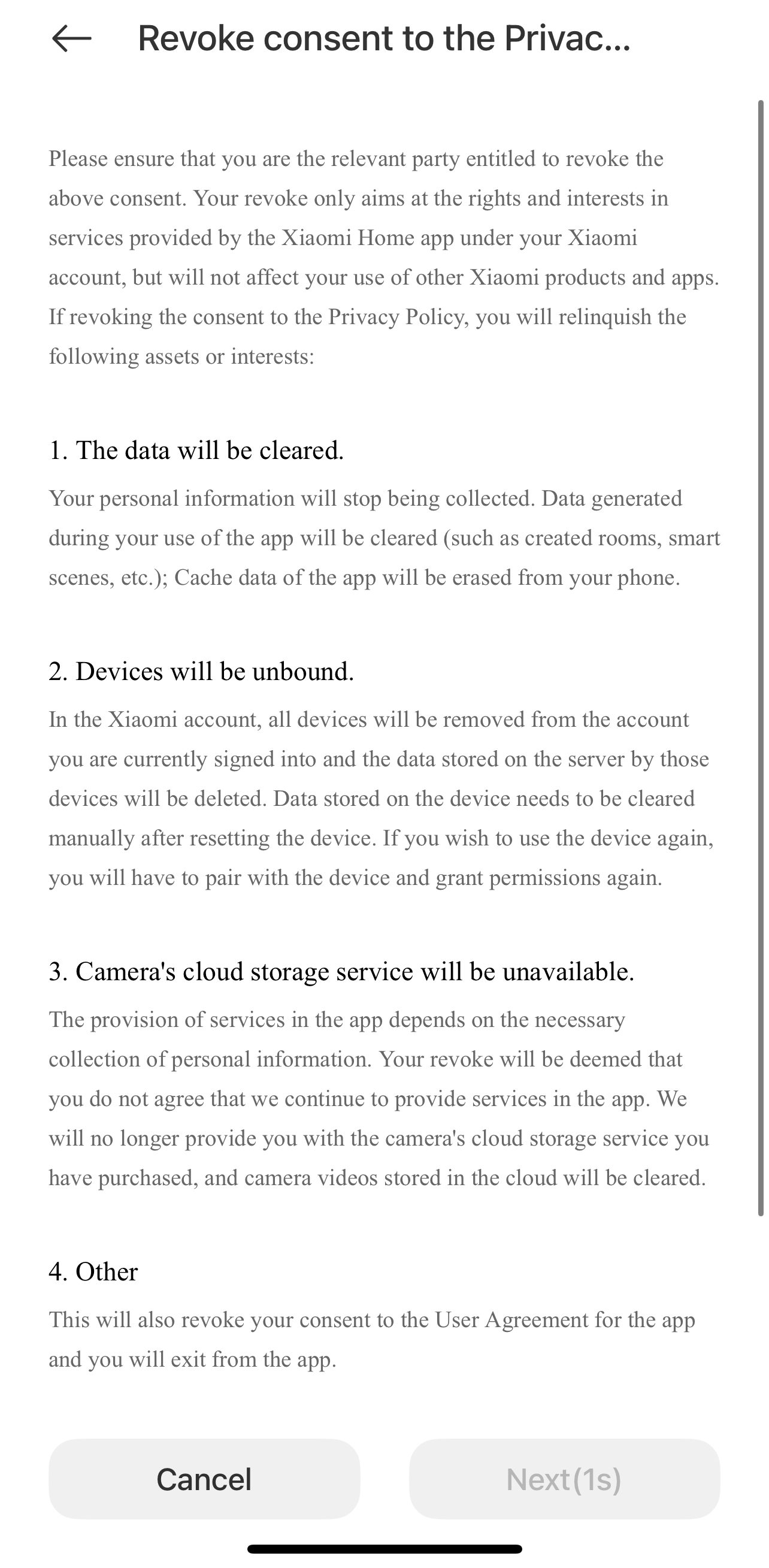}
        \caption{Revoking consent to privacy policy to delete specific user data (APP2).}
        \label{fig:4f}
    \end{subfigure}
    \hspace{0.02\textwidth}
    \begin{subfigure}[t]{0.22\textwidth}
        \centering
        \includegraphics[width=\textwidth]{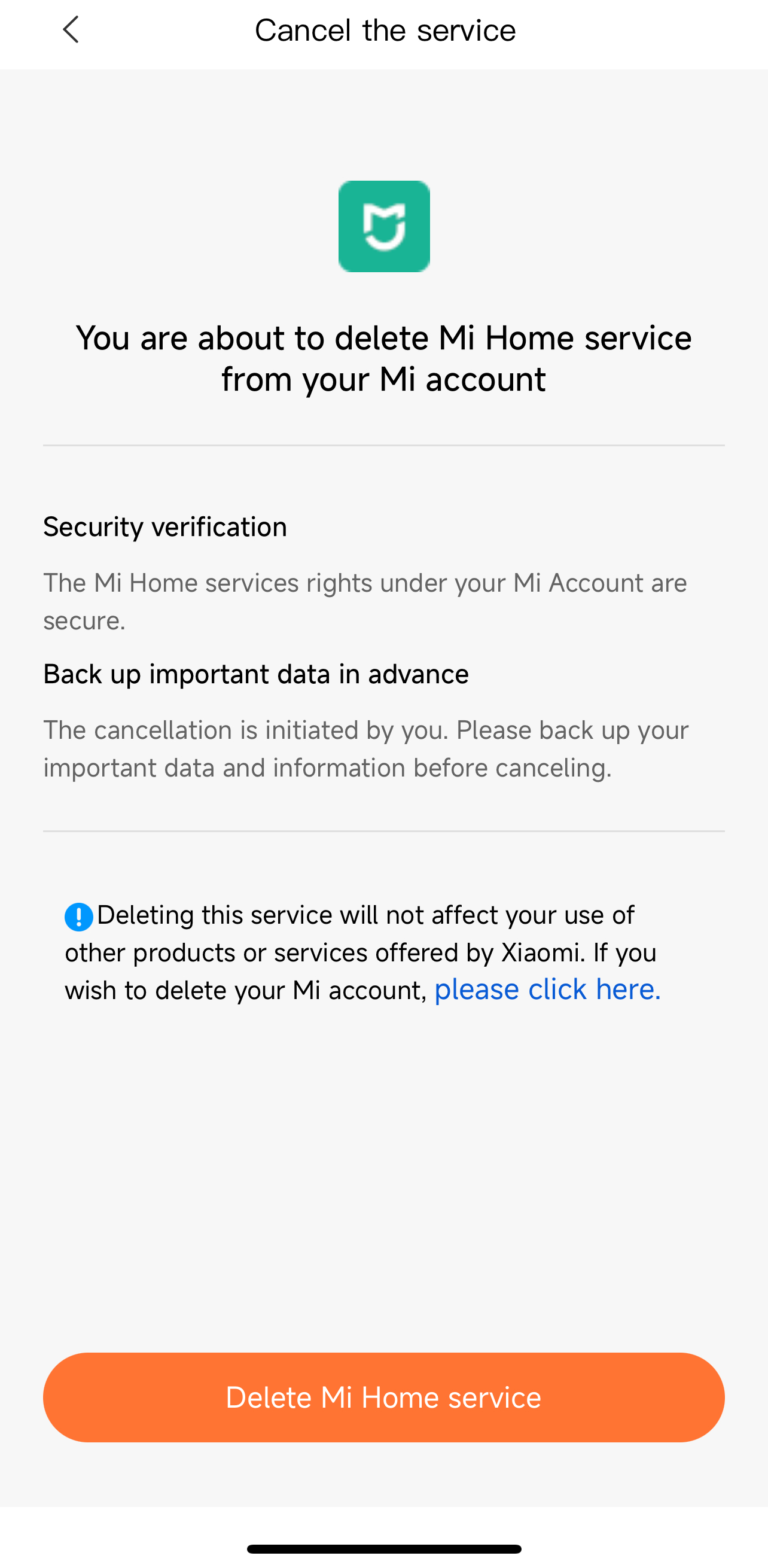}
        \caption{Option to delete account and erase all associated data (APP2).}
        \label{fig:4g}
    \end{subfigure}
    \caption{Examples of app deletion mechanisms.}
    \label{fig:data_deletion}
\end{figure*}

\begin{figure*}[htbp]
    \centering
    \hspace{0.02\textwidth}
    \begin{subfigure}[t]{0.22\textwidth}
        \centering
        \includegraphics[width=\textwidth]{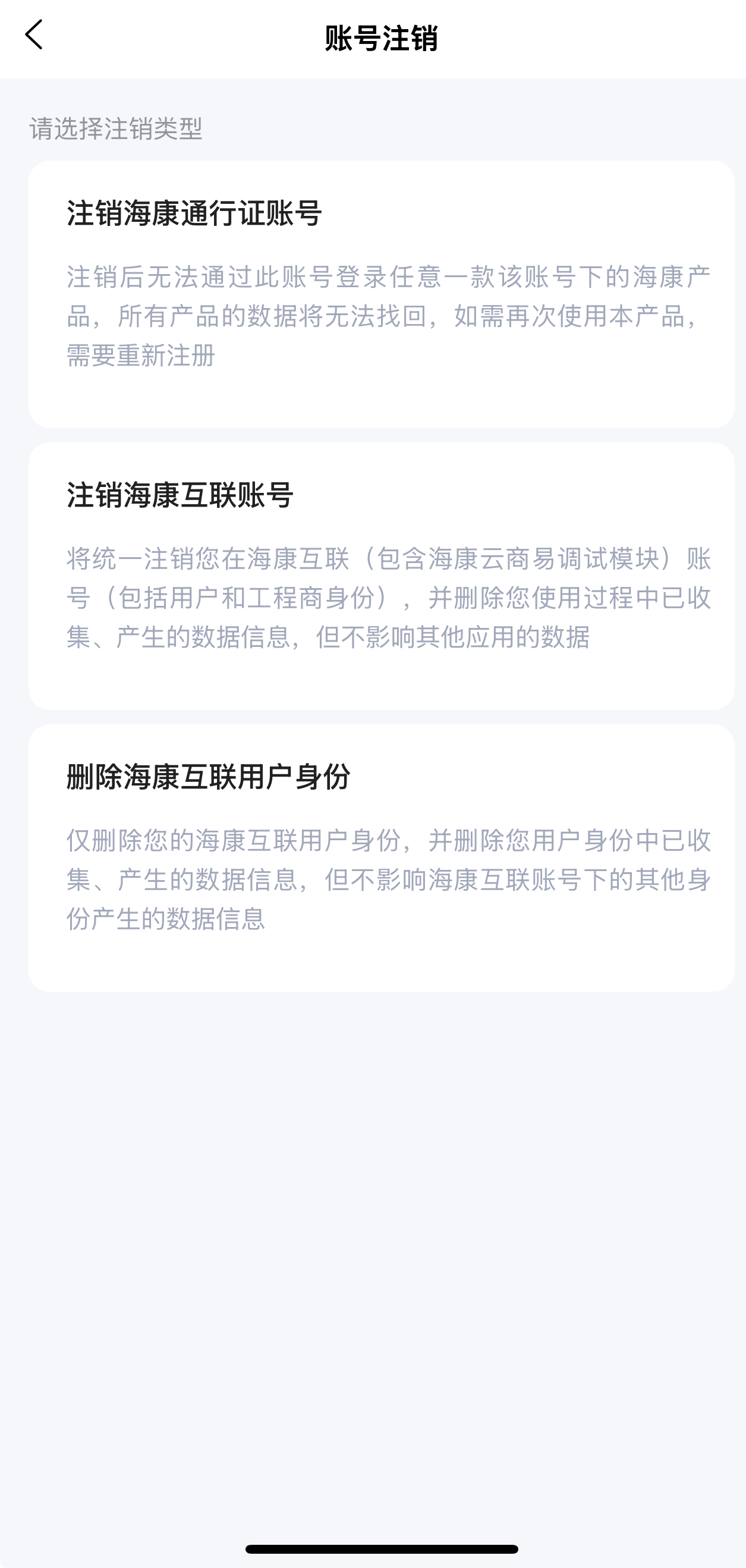}
        \caption{Deleting an app account managing different device types and providing multiple services would disable all linked services (APP7).}
        \label{fig:4h}
    \end{subfigure}
    \begin{subfigure}[t]{0.22\textwidth}
        \centering
        \includegraphics[width=\textwidth]{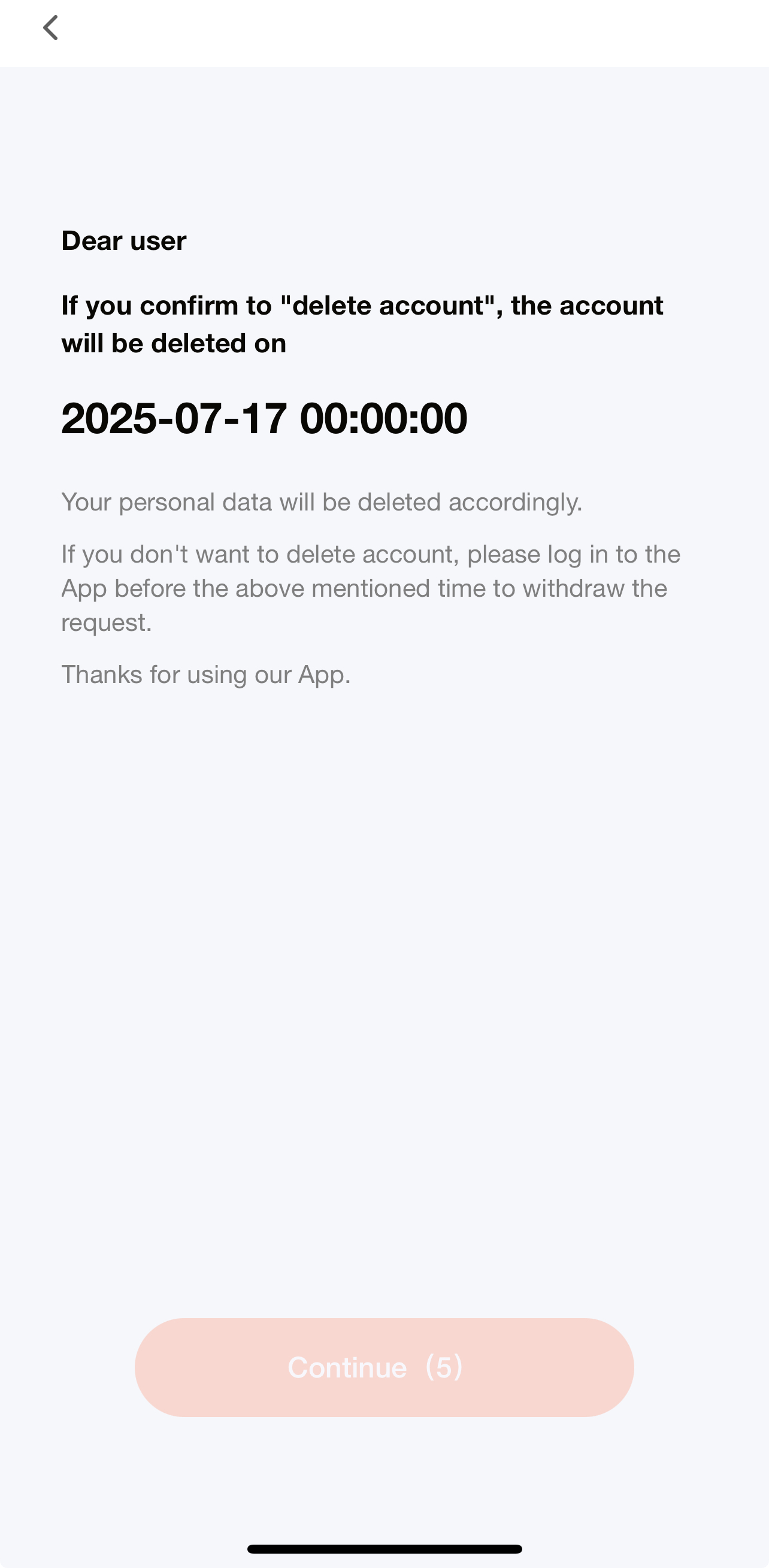}
        \caption{Repetitive options claiming to delete data (APP21).}
        \label{fig:4k}
    \end{subfigure}
    \hspace{0.02\textwidth}
    \begin{subfigure}[t]{0.22\textwidth}
        \centering
        \includegraphics[width=\textwidth]{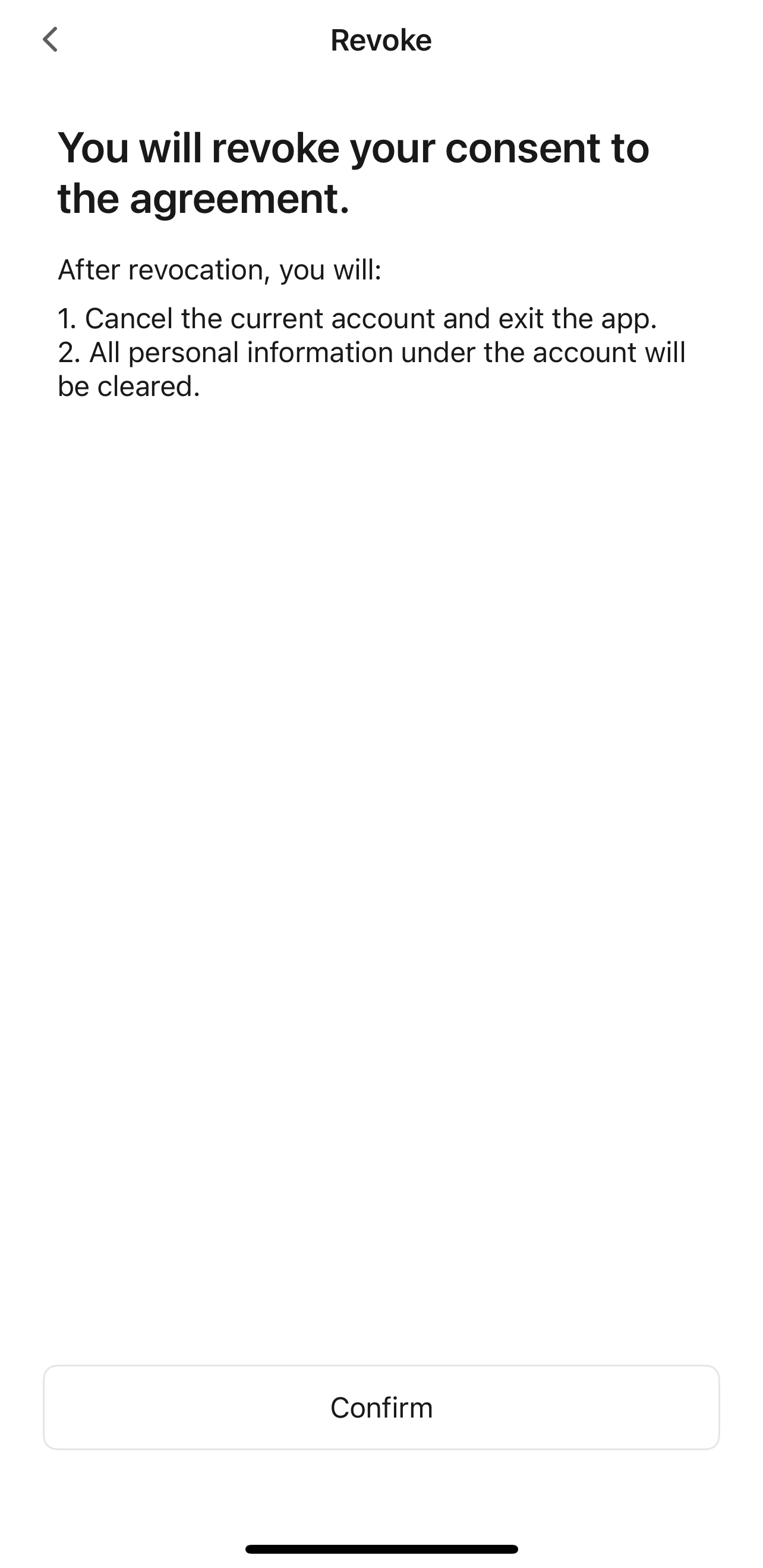}
        \caption{Unclear distinction between account deletion and consent revocation (APP21).}
        \label{fig:4l}
    \end{subfigure}
    \hspace{0.02\textwidth}
    \begin{subfigure}[t]{0.22\textwidth}
        \centering
        \includegraphics[width=\textwidth]{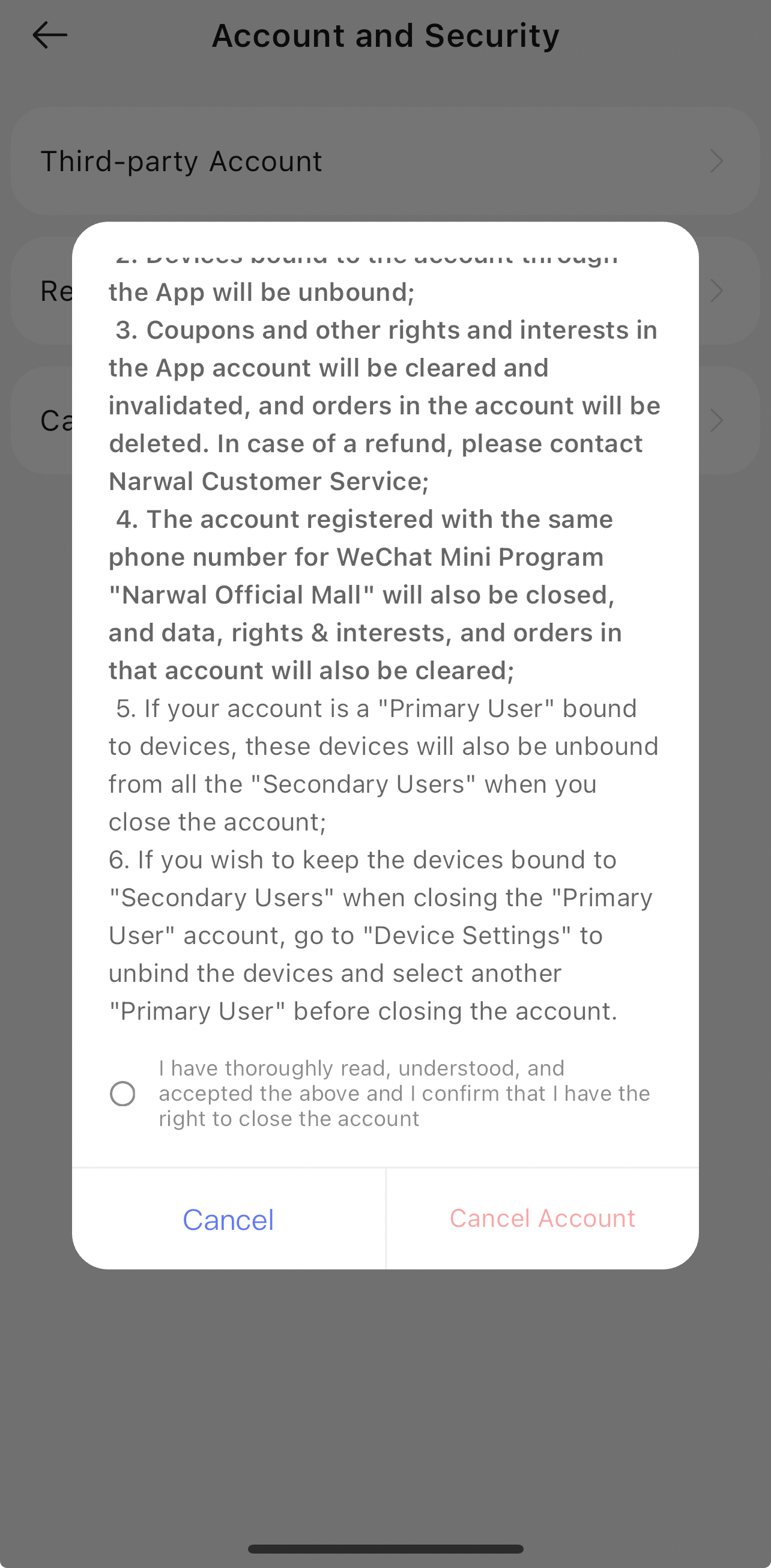}
        \caption{Role-based deletion rules lacked transparency (APP37).}
        \label{fig:4m}
    \end{subfigure}
    \\
    \begin{subfigure}[t]{0.22\textwidth}
        \centering
        \includegraphics[width=\textwidth]{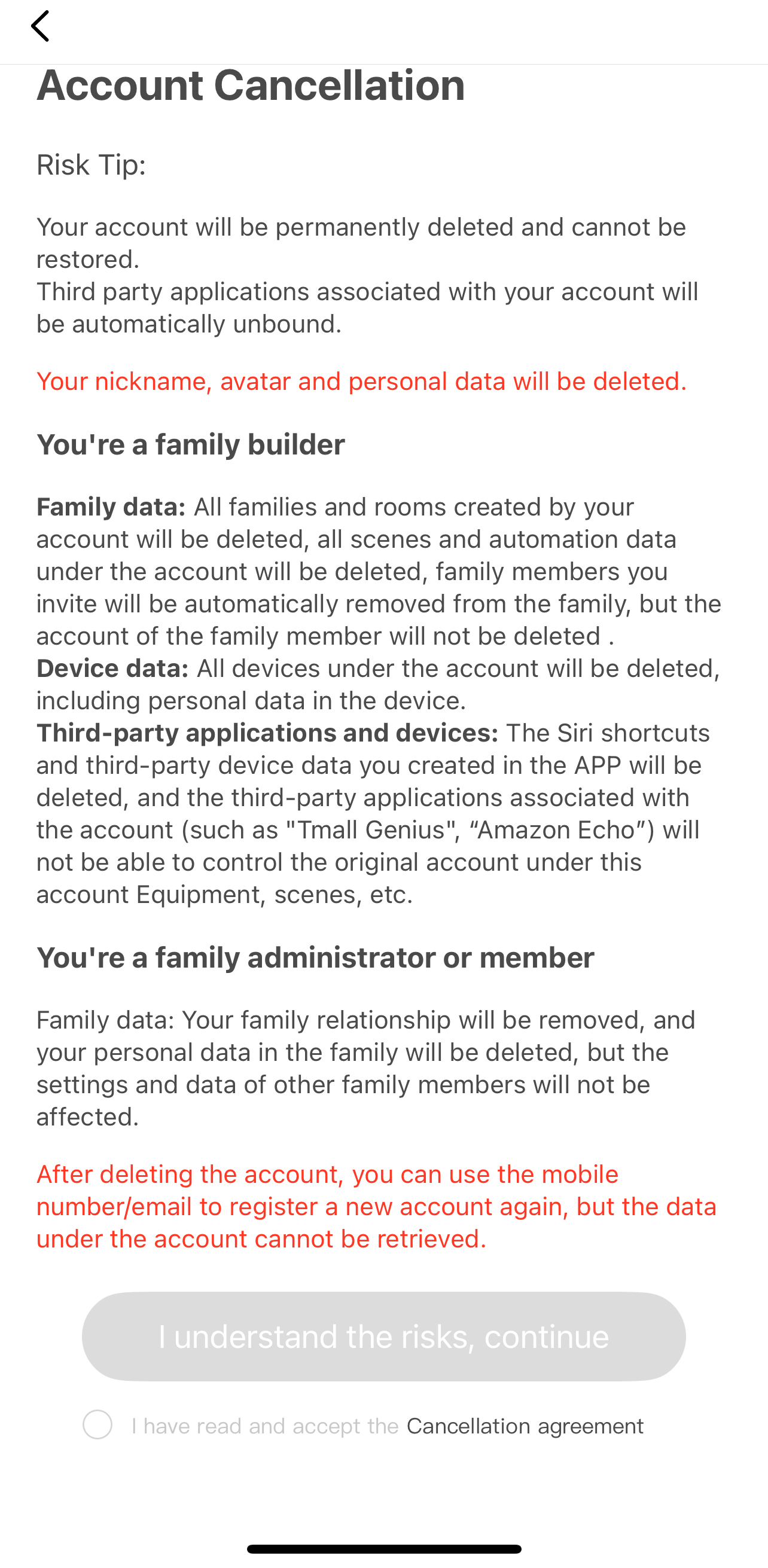}
        \caption{Non-transparent deletion across different user categories (APP42).}
        \label{fig:4n}
    \end{subfigure}
    \hspace{0.02\textwidth}
    \begin{subfigure}[t]{0.22\textwidth}
        \centering
        \includegraphics[width=\textwidth]{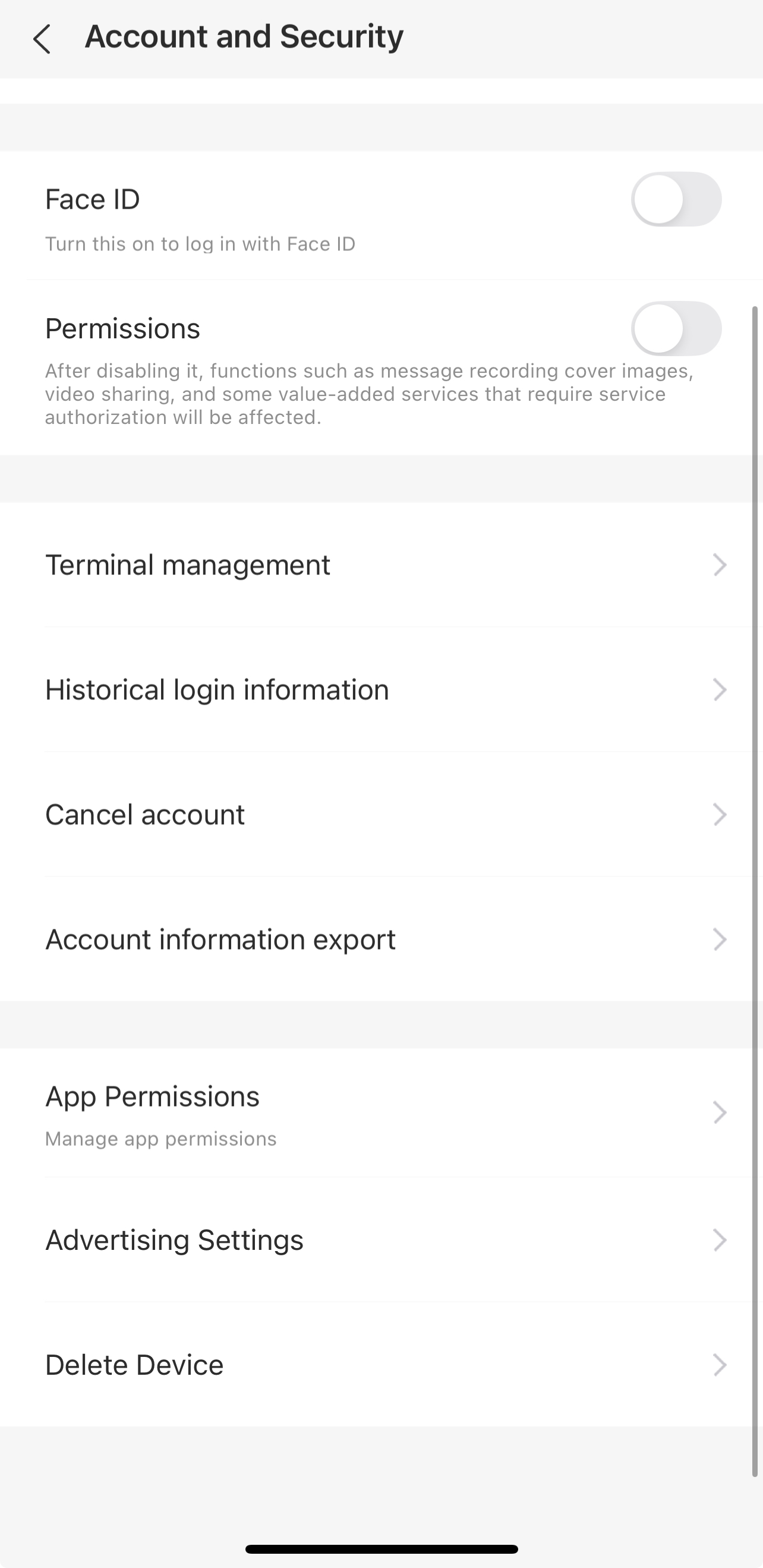}
        \caption{Connected devices remained active even after account deletion (APP44).}
        \label{fig:4o}
    \end{subfigure}
    \caption{Examples of usability and transparency challenges in app account deletion and consent flows.}
    \label{fig:multiple_deletion}
\end{figure*}%%

\subsubsection{App Data Deletion} \label{UX_data_deletion}
All apps provided data deletion mechanisms, typically located within user settings or the profile screen and easily accessible. Most apps clearly communicated the consequences of deletion, such as permanence and irreversibility, through on-screen warning pop-ups, and the actual outcomes generally aligned with these messages. Users were usually required to authenticate by entering their registered mobile number and corresponding password before proceeding. From our inspection of deletion practices, we identified three overarching patterns, differing in terms of data retention and recoverability:

\textbf{Full deletion option.} 35/49 apps provided a single deletion option that triggered full and irreversible removal of all user data (see Fig.~\ref{fig:4d}). Once deleted, the account could no longer access services, and all associated data became unrecoverable, including personal profiles and login data, home network settings, connected device information, cloud-generated personal data, third-party service data, and digital assets such as transactions, coupons, and remaining subscriptions.

\textbf{Separate options for data deletion and account deletion.} Four apps implemented a mechanism for revoking consent to the app privacy policy for all registered users (see Fig.~\ref{fig:4f}), which cleared specific data such as personal information, app cache, and information about connected devices. Additionally, these apps provided a separate account deletion option for users who wished to erase all data completely (see Fig.~\ref{fig:4g}).

\textbf{Multiple deletion options.} Some companies or developers designed their account systems to integrate a wider range of smart home devices and services, resulting in app-based ecosystems (see Fig.~\ref{fig:4h}). Users could delete either a specific app account or a central ecosystem account, with the latter disabling access to all related services and apps. We also identified three deceptive patterns related to ambiguous deletion practices: APP21 featured a repetitive and unclear process, where both the ``Delete Account'' and ``Revoke Consent'' options claimed to erase personal data (see Fig.~\ref{fig:4k}, Fig.~\ref{fig:4l}); APP37 and APP42 employed inconsistent and non-transparent mechanisms for different user roles, such as ``primary/secondary user'' and ``family builder/family administrator'' (see Fig.~\ref{fig:4m}, Fig.~\ref{fig:4n}); and in APP44, account and personal data deletion required separate unbinding of devices (see Fig.~\ref{fig:4o}). Regarding bystander data deletion, no app in our analysis provided mechanisms directly accessible to bystanders. Consequently, bystander privacy management depended entirely on account holders’ discretion, significantly limiting their control over personal data removal.%%

\begin{shaded}
\textbf{RQ2:} Most apps incorporated standard mechanisms for account creation, permission requests, and data deletion, often complying with legal requirements such as real-name registration and mandatory consent for privacy policies. While some apps provided intuitive, centralized privacy settings with granular control, many buried these functions or lacked clear interfaces altogether. Importantly, bystander privacy remained poorly supported across all UX stages, regardless of the app. No app offered consent or notification mechanisms for non-users when handling registration, permission requests, or data deletion. Bystanders, such as family members or visitors, lacked visibility and agency over how their data was captured or deleted, as all controls resided with primary account holders. This imbalance reveals a critical gap: despite the growing sophistication of the smart home app ecosystem, current UX/UI designs prioritized ensuring that apps complied with legal and regulatory requirements and seamless integration into company ecosystems over implementing inclusive, bystander-aware privacy protections.
\end{shaded}%%%

\subsection{Traceability Analysis} \label{traceability_analyse}
We conducted an in-depth traceability analysis to systematically map explicit privacy policy provisions against actual UX/UI implementations across the 49 apps selected in our study (see Table~\ref{tab:traceability_table}). We synthesized these findings using a comprehensive traceability matrix, highlighting compliance gaps and areas of disconnect, as well as identifying critical opportunities to enhance transparency, usability, and user trust. This analysis cross-references privacy policy commitments with observed UX/UI practices across four key dimensions:
    %\item \textbf{Data Sharing} captures user-initiated sharing with other users (such as family members) and provider-initiated disclosures to third parties (e.g., SDK services) or law enforcement (e.g., public security requests). Privacy policies explicitly describing these practices were cross-referenced with corresponding UI features enabling users to control these actions; 
    %\item \textbf{Data Control and Management} covers UI implementations enabling users to exercise control over their data, such as permission settings, data deletion, retention and portability, personal information review or modification, opt-out mechanisms for marketing, account deletion (both app-based and ecosystem-based), age restrictions for minors, and bystander data management (live-in, visiting, and uninvolved bystanders~\cite{saqib2025bystander}). The evaluation assessed the completeness of these UI features in fulfilling explicit privacy policy promises, applying the same four-level evaluation framework; 
    %\item \textbf{Security Measures} assesses explicit mentions of technical data encryption methods (e.g., anonymization) within privacy policies and verifies their visible implementation through UI practices or communicated functionalities. 
%\end{enumerate}

\begin{table*}[htbp]
\resizebox{1\textwidth}{!}{%
\begin{tabular}{l|cc|cccccc|ccc|cccccccccc|c}
\toprule
\multirow{4}{*}{\textbf{ID}} 
& \multicolumn{8}{c|}{\textbf{Data Collection}} 
& \multicolumn{3}{c|}{\textbf{Data Sharing by}} 
& \multicolumn{10}{c|}{\textbf{Data Control and Management}} 
& \multicolumn{1}{c}{\textbf{Security Measures}} \\
\cmidrule(lr){2-9} \cmidrule(lr){10-12} \cmidrule(lr){13-22} \cmidrule(lr){23-23}
& \multicolumn{2}{c|}{\textbf{Direct Collection}} 
& \multicolumn{6}{c|}{\textbf{Indirect Collection}}
& \multicolumn{1}{c|}{\textbf{Users}}
& \multicolumn{2}{c|}{\textbf{App Providers}}
& \multicolumn{10}{c|}{} 
& \multicolumn{1}{c}{} \\
\cmidrule(lr){2-3} \cmidrule(lr){4-9} \cmidrule(lr){10-10} \cmidrule(lr){11-12} 
& \rotatebox{90}{PII/Registration Info} 
& \rotatebox{90}{Optional Info} 
& \rotatebox{90}{Device Identifiers} 
& \rotatebox{90}{Location Data} 
& \rotatebox{90}{Multimedia Data} 
& \rotatebox{90}{App Logs/Interactions} 
& \rotatebox{90}{Financial Data} 
& \rotatebox{90}{Passive Info} 
& \rotatebox{90}{With Other (Live-in) Users} 
& \rotatebox{90}{With Third-party Services} 
& \rotatebox{90}{With Law Enforcement} 
& \rotatebox{90}{User Privacy Controls} 
& \rotatebox{90}{Data Deletion} 
& \rotatebox{90}{Data Retention} 
& \rotatebox{90}{Reviewing  Account Info} 
& \rotatebox{90}{Personalization/Marketing} 
& \rotatebox{90}{Account Deletion} 
& \rotatebox{90}{Age Restrictions} 
& \rotatebox{90}{Bystander Priv Mgmt. (Live-in)} 
& \rotatebox{90}{Bystander Priv. Mgmt. (Visiting)} 
& \rotatebox{90}{Bystander Priv. mgmt. (Uninvolved)} 
& \rotatebox{90}{Data Encryption} \\
\midrule
APP1 & \complete & \complete & \complete & \complete & \complete & \complete & \complete & \complete & \notapplicable & \complete & \broken & \notcomplete & \notcomplete & \notcomplete & \complete & \complete & \complete & \complete & \notapplicable & \notapplicable & \notapplicable & \complete \\
APP2 & \complete & \complete & \complete & \complete & \complete & \complete & \complete & \complete & \complete & \complete & \broken & \complete & \complete & \notcomplete & \complete & \complete & \complete & \complete & \complete & \notapplicable & \notapplicable & \complete \\
APP3 & \complete & \complete & \complete & \complete & \complete & \complete & \complete & \complete & \notapplicable & \complete & \broken & \complete & \notcomplete & \notcomplete & \complete & \complete & \complete & \complete & \notapplicable & \notapplicable & \notapplicable & \complete \\
APP4 & \complete & \complete & \complete & \complete & \complete & \complete & \complete & \complete & \notapplicable & \complete & \broken & \notcomplete & \complete & \notcomplete & \complete & \complete & \complete & \complete & \notapplicable & \notapplicable & \notapplicable & \complete \\
APP5 & \complete & \complete & \complete & \complete & \complete & \complete & \complete & \complete & \notcomplete & \complete & \broken & \notcomplete & \notcomplete & \notcomplete & \complete & \complete & \complete & \complete & \notcomplete & \notapplicable & \notapplicable & \complete \\
APP6 & \complete & \complete & \complete & \complete & \complete & \complete & \complete & \complete & \notapplicable & \complete & \broken & \notcomplete & \notcomplete & \notcomplete & \complete & \notapplicable & \complete & \complete & \notapplicable & \notapplicable & \notapplicable & \complete \\
APP7 & \complete & \complete & \complete & \complete & \complete & \complete & \complete & \complete & \notapplicable & \complete & \broken & \notapplicable & \complete & \complete & \complete & \complete & \complete & \complete & \notapplicable & \notapplicable & \notapplicable & \complete \\
APP8 & \complete & \complete & \complete & \complete & \complete & \complete & \complete & \complete & \notcomplete & \complete & \broken & \complete & \complete & \complete & \complete & \complete & \complete & \complete & \notcomplete & \notapplicable & \notapplicable & \complete \\
APP9 & \complete & \complete & \complete & \complete & \complete & \complete & \complete & \complete & \notcomplete & \complete & \broken & \complete & \notcomplete & \notcomplete & \complete & \complete & \complete & \complete & \notcomplete & \notapplicable & \notapplicable & \complete \\
APP10 & \complete & \complete & \complete & \complete & \complete & \complete & \complete & \complete & \notcomplete & \complete & \broken & \complete & \notcomplete & \notcomplete & \complete & \complete & \complete & \complete & \notcomplete & \notapplicable & \notapplicable & \complete\\
APP11 & \complete & \complete & \complete & \complete & \complete & \complete & \complete & \complete & \complete & \complete & \broken & \notcomplete & \notcomplete & \notcomplete & \complete & \complete & \complete & \complete & \complete & \notapplicable & \notapplicable & \complete\\
APP12 & \complete & \complete & \complete & \complete & \complete & \complete & \notapplicable & \complete & \notcomplete & \complete & \broken & \notcomplete & \notcomplete & \notcomplete & \complete & \notapplicable & \complete & \complete & \notcomplete & \notapplicable & \notapplicable & \complete\\
APP13 & \complete & \complete & \complete & \complete & \complete & \complete & \complete & \complete & \notapplicable & \complete & \broken & \notcomplete & \complete & \notcomplete & \complete & \notapplicable & \complete & \complete & \notapplicable & \notapplicable & \notapplicable & \complete \\
APP14 & \complete & \complete & \complete & \complete & \complete & \complete & \complete & \complete & \notapplicable & \complete & \broken & \complete & \complete & \notcomplete & \complete & \notapplicable & \complete & \complete & \notapplicable & \notapplicable & \notapplicable & \complete \\
APP15 & \complete & \complete & \complete & \complete & \complete & \complete & \complete & \complete & \notcomplete & \complete & \broken & \complete & \complete & \notcomplete & \complete & \notapplicable & \complete & \complete & \notcomplete & \notapplicable & \notapplicable & \complete\\
APP16 & \complete & \complete & \complete & \complete & \complete & \complete & \complete & \complete & \notapplicable & \complete & \broken & \notcomplete & \notcomplete & \notcomplete & \complete & \complete & \complete & \complete & \notapplicable & \notapplicable & \notapplicable & \complete\\
APP17 & \complete & \complete & \complete & \complete & \complete & \complete & \complete & \complete & \notcomplete & \complete & \broken & \notcomplete & \notcomplete & \notcomplete & \complete & \complete & \complete & \complete & \notcomplete & \notapplicable & \notapplicable & \complete\\
APP18 & \complete & \complete & \complete & \complete & \complete & \complete & \complete & \complete & \notapplicable & \complete & \broken & \complete & \complete & \notcomplete & \complete & \notapplicable & \complete & \complete & \notapplicable & \notapplicable & \notapplicable & \complete\\
APP19 & \complete & \complete & \complete & \complete & \complete & \complete & \complete & \complete & \notapplicable & \complete & \broken & \complete & \notcomplete & \complete & \complete & \complete & \complete & \complete & \notapplicable & \notapplicable & \notapplicable & \complete\\
APP20 & \complete & \complete & \complete & \complete & \complete & \complete & \notapplicable & \complete & \notapplicable & \complete & \broken & \notcomplete & \complete & \notcomplete & \complete & \notapplicable & \complete & \complete & \notapplicable & \notapplicable & \notapplicable & \complete\\
APP21 & \complete & \complete & \complete & \complete & \complete & \complete & \complete & \complete & \notapplicable & \complete & \broken & \complete & \complete & \complete & \complete & \complete & \complete & \complete & \notapplicable & \notapplicable & \notapplicable & \complete \\
APP22 & \complete & \complete & \complete & \complete & \complete & \complete & \notapplicable & \complete & \complete & \complete & \broken & \notcomplete & \notcomplete & \notcomplete & \complete & \notapplicable & \complete & \complete & \complete & \notapplicable & \notapplicable & \complete \\
APP23 & \complete & \complete & \complete & \complete & \complete & \complete & \complete & \complete & \notapplicable & \complete & \broken & \notcomplete & \notcomplete & \notcomplete & \complete & \complete & \complete & \complete & \notapplicable & \notapplicable & \notapplicable & \complete \\
APP24 & \complete & \complete & \complete & \complete & \complete & \complete & \complete & \complete & \notapplicable & \complete & \broken & \notcomplete & \notcomplete & \notcomplete & \complete & \notapplicable & \complete & \complete & \notapplicable & \notapplicable & \notapplicable & \complete \\
APP25 & \complete & \complete & \complete & \complete & \complete & \complete & \complete & \complete & \notcomplete & \complete & \broken & \notcomplete & \complete & \notcomplete & \complete & \complete & \complete & \complete & \notcomplete & \notapplicable & \notapplicable & \complete \\
APP26 & \complete & \complete & \complete & \complete & \complete & \complete & \complete & \complete & \notapplicable & \complete & \broken & \complete & \complete & \notcomplete & \complete & \complete & \complete & \complete & \notapplicable & \notapplicable & \notapplicable & \complete \\
APP27 & \complete & \complete & \complete & \complete & \complete & \complete & \complete & \complete & \complete & \complete & \broken & \notcomplete & \notcomplete & \notcomplete & \complete & \complete & \complete & \complete & \complete & \notapplicable & \notapplicable & \complete \\
APP28 & \complete & \complete & \complete & \complete & \complete & \complete & \notapplicable & \complete & \notcomplete & \complete & \broken & \notcomplete & \notcomplete & \notcomplete & \complete & \complete & \complete & \complete & \notcomplete & \notapplicable & \notapplicable & \complete \\
APP29 & \complete & \complete & \complete & \complete & \complete & \complete & \complete & \complete & \notcomplete & \complete & \broken & \notcomplete & \notcomplete & \notcomplete & \complete & \complete & \complete & \complete & \notcomplete & \notapplicable & \notapplicable & \complete \\
APP30 & \complete & \complete & \complete & \complete & \complete & \complete & \complete & \complete & \complete & \complete & \broken & \complete & \notcomplete & \complete & \complete & \complete & \complete & \complete & \complete & \notapplicable & \notapplicable & \complete \\
APP31 & \complete & \complete & \complete & \complete & \complete & \complete & \complete & \complete & \notapplicable & \complete & \broken & \complete & \notcomplete & \notcomplete & \complete & \complete & \complete & \complete & \notapplicable & \notapplicable & \notapplicable & \complete \\
APP32 & \complete & \complete & \complete & \complete & \complete & \complete & \complete & \complete & \complete & \complete & \broken & \complete & \notcomplete & \notcomplete & \complete & \complete & \complete & \complete & \complete & \notapplicable & \notapplicable & \complete \\
APP33 & \complete & \complete & \complete & \complete & \complete & \complete & \complete & \complete & \notapplicable & \complete & \broken & \complete & \complete & \notcomplete & \complete & \broken & \complete & \complete & \notapplicable & \notapplicable & \notapplicable & \complete \\
APP34 & \complete & \complete & \complete & \complete & \complete & \complete & \complete & \complete & \notcomplete & \complete & \broken & \complete & \notcomplete & \complete & \complete & \complete & \complete & \complete & \notcomplete & \notapplicable & \notapplicable & \complete \\
APP35 & \complete & \complete & \complete & \complete & \complete & \complete & \complete & \complete & \notapplicable & \complete & \broken & \notcomplete & \notcomplete & \notcomplete & \complete & \notapplicable & \complete & \complete & \notapplicable & \notapplicable & \notapplicable & \complete \\
APP36 & \complete & \complete & \complete & \complete & \complete & \complete & \notapplicable & \complete & \complete & \complete & \broken & \notcomplete & \notcomplete & \notcomplete & \complete & \notapplicable & \complete & \complete & \complete & \notapplicable & \notapplicable & \complete \\
APP37 & \complete & \complete & \complete & \complete & \complete & \complete & \complete & \complete & \notcomplete & \complete & \broken & \notcomplete & \notcomplete & \notcomplete & \complete & \notapplicable & \complete & \complete & \notcomplete & \notapplicable & \notapplicable & \complete \\
APP38 & \complete & \complete & \complete & \complete & \complete & \complete & \complete & \complete & \notapplicable & \complete & \broken & \notcomplete & \complete & \notcomplete & \complete & \broken & \complete & \complete & \notapplicable & \notapplicable & \notapplicable & \complete \\
APP39 & \complete & \complete & \complete & \complete & \complete & \complete & \notapplicable & \complete & \notcomplete & \complete & \broken & \notcomplete & \complete & \notcomplete & \complete & \notapplicable & \complete & \broken & \notcomplete & \notapplicable & \notapplicable & \complete \\
APP40 & \complete & \complete & \complete & \complete & \complete & \complete & \notapplicable & \complete & \notapplicable & \complete & \broken & \notcomplete & \complete & \notcomplete & \complete & \broken & \complete & \complete & \notapplicable & \notapplicable & \notapplicable & \complete \\
APP41 & \complete & \complete & \complete & \complete & \complete & \complete & \broken & \complete & \notapplicable & \complete & \broken & \notcomplete & \complete & \notcomplete & \complete & \broken & \complete & \broken & \notapplicable & \notapplicable & \notapplicable & \complete \\
APP42 & \complete & \complete & \complete & \complete & \complete & \complete & \notapplicable & \complete & \notapplicable & \complete & \broken & \notcomplete & \complete & \notcomplete & \complete & \notapplicable & \complete & \complete & \notapplicable & \notapplicable & \notapplicable & \complete \\
APP43 & \complete & \complete & \complete & \complete & \complete & \complete & \complete & \complete & \complete & \complete & \broken & \notcomplete & \notcomplete & \notcomplete & \complete & \complete & \complete & \complete & \complete & \notapplicable & \notapplicable & \complete \\
APP44 & \complete & \complete & \complete & \complete & \complete & \complete & \complete & \complete & \notapplicable & \complete & \broken & \notcomplete & \complete & \complete & \complete & \complete & \complete & \complete & \notapplicable & \notapplicable & \notapplicable & \complete \\
APP45 & \complete & \complete & \complete & \complete & \complete & \complete & \notapplicable & \complete & \notcomplete & \complete & \broken & \notcomplete & \complete & \notcomplete & \complete & \notapplicable & \complete & \complete & \notcomplete & \notapplicable & \notapplicable & \complete \\
APP46 & \complete & \complete & \complete & \complete & \complete & \complete & \notapplicable & \complete & \notapplicable & \complete & \broken & \notcomplete & \notcomplete & \notcomplete & \complete & \notapplicable & \complete & \complete & \notapplicable & \notapplicable & \notapplicable & \complete \\
APP47 & \complete & \complete & \complete & \complete & \complete & \complete & \complete & \complete & \complete & \complete & \broken & \complete & \complete & \notcomplete & \complete & \complete & \complete & \complete & \complete & \notapplicable & \notapplicable & \complete \\
APP48 & \complete & \complete & \complete & \complete & \complete & \complete & \complete & \complete & \complete & \complete & \broken & \complete & \notcomplete & \complete & \complete & \notapplicable & \complete & \complete & \complete & \notapplicable & \notapplicable & \complete \\
APP49 & \complete & \complete & \complete & \complete & \complete & \complete & \complete & \complete & \notapplicable & \complete & \broken & \complete & \notcomplete & \notcomplete & \complete & \complete & \complete & \complete & \notapplicable & \notapplicable & \notapplicable & \complete \\
\bottomrule
\end{tabular}}
\caption{Traceability analysis matrix for Chinese smart home apps ($n=49$).}
\label{tab:traceability_table}
\end{table*}%%%

\textbf{Data Collection.} All Direct Collection items, including PII and registration details (e.g., real names, ID numbers) and optional information (e.g., avatars, nicknames), were generally evaluated as having \textit{Complete} traceability across the 49 apps. For Indirect Collection, elements such as device identifiers (e.g., IMEI, MAC addresses), location data (e.g., Wi-Fi SSIDs), multimedia and sensor access (e.g., camera, microphone), app logs, interaction data, and passively gathered information (e.g., cookies) also showed \textit{Complete} traceability---being clearly disclosed in policies and consistently implemented in UI settings. However, for financial data (e.g., bank account details, third-party payment methods), ten apps were marked \textit{Not Applicable}---as their policies omitted financial data collection and their UIs lacked relevant controls. Notably, APP41 was rated as having \textit{Broken} traceability: although its policy stated that it collected financial information such as order history and payment methods, the app interface offered no means to review, manage, or delete such data. This absence of UI-level control undermined transparency and user agency over sensitive financial information.%%%

\textbf{Data Sharing.} Data shared with third-party entities, such as advertisers or SDK providers, was clearly communicated in both policies and app UIs across all 49 apps, hence the apps were generally evaluated as having \textit{Complete} traceability. Although all apps referenced compliance with legal requirements under Art. 35, CSL regarding data sharing with law enforcement or public security authorities, none provided UI features---such as notifications or visual indicators in in-app settings---to inform users about this data-sharing mechanism. Consequently, all apps were evaluated as having \textit{Broken} traceability in this context. Regarding user-initiated sharing, particularly with live-in bystanders such as household members, 25/49 apps were rated as \textit{Not Applicable}, as their policies and UIs lacked explicit features or mechanisms enabling user-to-user data sharing. Additionally, 14/49 apps were evaluated as having \textit{Partial} traceability, primarily due to the presence of some user data-sharing settings without clear policy descriptions or guidance on consent mechanisms. For example, APP34 allowed users to invite family members via third-party platforms like WeChat but did not adequately explain the associated privacy implications or available management controls in its policy documents.

\textbf{Data Control and Management.} In compliance with PIPL (Art. 47, item 3), all apps provided mechanisms for account deletion and enabled reviewing and modifying personal information (e.g., PII and account details), resulting in \textit{Complete} traceability. For managing personalized and marketing content, 28/49 apps demonstrated consistency between policy descriptions and UI implementations, achieving \textit{Complete} traceability. However, four apps were evaluated as having \textit{Broken} traceability because their policies described user controls for personalized or marketing content, but the UI lacked corresponding settings. For instance, APP13's policy clearly described such controls, yet its UI lacked identifiable options, rendering them practically inaccessible. Moreover, only APP39 and APP41 failed to address age restrictions, receiving \textit{Broken} traceability; specifically, APP39 did not specify age-related protections in either its policy or UI, a significant omission under the child protection provisions of PIPL and PCPCPI.

We also evaluated bystander data handling practices, categorizing bystanders into live-in, visiting, and uninvolved types~\cite{saqib2025bystander}. All results related to live-in bystanders mirrored findings from the user-sharing category, as permissions and access controls for household members also applied to live-in bystanders. These settings allowed primary users to invite, share, or manage access---potentially extending to domestic workers---though no app explicitly referenced such groups. In contrast, for visiting and uninvolved bystanders, none of the apps provided explicit mentions or UI practices, resulting in \textit{Not Applicable} traceability across all 49 apps. This absence highlights a major gap in current privacy protections, leaving these bystanders without clear rights or control in Chinese smart home app contexts.%%

\textbf{Security and Privacy Measures.} All apps were categorized as having \textit{Complete} traceability, as they consistently outlined robust technical and organizational safeguards, such as data encryption, anonymization, access controls, and incident response procedures, that were reflected in the corresponding app interfaces, in adherence to PIPL requirements (Art. 51). These measures were typically accompanied by visible security and privacy settings or notifications, reinforcing alignment between policy commitments and actual UI practices.

\subsubsection{Privacy Labels} \label{result_privacy_label}
We also analyzed the privacy labels of the apps as displayed on the Apple App Store (see Table~\ref{tab:privacy_labels}) and compared them with our traceability results. Our findings revealed widespread inconsistencies and underreporting across nearly all apps, indicating that these privacy labels often failed to accurately reflect actual data practices.

Specifically, 26/49 apps claimed in their privacy labels that they did not collect any data used to track users, while their privacy policies acknowledged using third-party SDKs for behavior monitoring and analytics, revealing a significant traceability mismatch. According to Apple, data used to track users includes any information linked across third-party apps or websites (typically via embedded SDKs) for targeted advertising or measuring ad performance. Notably, all 49 apps disclosed sharing data with SDK providers (e.g., for analytics, payments, or push notifications) in their privacy policies, suggesting these SDKs might enable cross-app tracking, even if not explicitly labeled as advertising-related, thereby contradicting App Store disclosures. Furthermore, although our traceability analysis confirmed that all apps collected direct PII, such as phone numbers with real-name verification for account registration and authentication, 23/49 apps claimed in their privacy labels that they did not collect any data linked to the user. Apple defines such data as tied to a user's identity, including account information, device identifiers, or contact details.\footnote{https://support.apple.com/zh-cn/102399} This discrepancy suggests either misunderstanding or intentional under-reporting.

In addition, six apps (APP7, APP28, APP33, APP38, APP45, APP48) explicitly declared in their privacy labels that they collected no data at all, yet our analysis found otherwise. The privacy policies of all these apps disclosed at least some data collection and/or sharing data with third-party SDKs (e.g., granting access to user location, camera, or microphone, or mandating account registration), contradicting the ``no data'' labels. Thus, the mismatch was primarily between the labels and policy/UI evidence rather than undisclosed processing. Such misreporting could undermine the credibility of privacy labels and mislead users seeking to make informed decisions.%

\begin{shaded}
\textbf{RQ3:} Our traceability analysis reveals that while most apps disclosed data collection, sharing, and security measures in line with PIPL requirements, many failed to reflect these provisions in their interfaces, particularly regarding user data controls and consent mechanisms. Importantly, none of the apps offered UI features to inform users about data sharing with authorities. Notably, bystander privacy remained largely unaddressed: live-in bystanders, such as family members, were only \emph{partially} covered through shared user settings, while visiting and uninvolved bystanders lacked any explicit protections across all 49 apps. Finally, we identified widespread discrepancies between the Apple App Store privacy labels and actual practices, including underreporting of tracking and personal data collection, thereby undermining transparency and eroding user trust.
\end{shaded}%%%%%%%
\section{Discussion}
\subsection{Lack of Bystander Privacy Considerations in Chinese Smart Home Apps}

Drawing on our findings from the privacy policy analysis (RQ1, \S\ref{privacy_policy}), UX/UI analysis (RQ2, \S\ref{ux_analyse}), and traceability analysis (RQ3, \S\ref{traceability_analyse}), our study reveals a significant gap in how bystander privacy is addressed. While Chinese smart home apps generally provide relatively mature mechanisms for protecting the privacy and security of primary users, bystander privacy remains systematically under-specified across all layers of the ecosystem---particularly for visiting or otherwise uninvolved individuals.

Although some apps (19/49) offered data-sharing settings for family members (e.g., voiceprint recognition or health tracking for cohabitants), none explicitly addressed the privacy of visiting bystanders (e.g., guests or short-term visitors) or uninvolved bystanders (e.g., neighbors or passersby) in either their policy documents (RQ1) or their app UIs (RQ2). The few bystander-relevant settings that did exist (RQ2) were tightly coupled with family-sharing features. For instance, several apps allowed users to input family health information and authorize monitoring functions such as heart rate or sleep tracking [APP10], yet provided no guidance on informed consent or data boundaries for cohabitants whose data might be collected. Similarly, APP4 enabled personalized voice services based on audio captured from all household members but offered no privacy warnings or consent options for individuals whose voices might be incidentally recorded. Together, these findings suggest that bystander privacy is treated as an exceptional or peripheral concern rather than a core design consideration. Smart home technologies inherently extend data collection beyond the primary user---often capturing information about individuals who have limited awareness of or control over these systems. Consequently, the lack of explicit bystander protections represents a substantive gap in current smart home privacy practices.%%

As shown in the traceability matrix (see Table~\ref{tab:traceability_table}, RQ3), all apps were rated as \textit{Not Applicable} in bystander-related categories. This rating reflects not only the absence of bystander-related privacy features in app UIs or their documentation in privacy policies, but also a broader lack of conceptual recognition of bystander privacy. These findings align with He et al.’s analysis of privacy considerations in Chinese smart home product design and development~\cite{he2025privacy}, which shows that product teams tend to prioritize legal compliance and focus primarily on the privacy of device owners or primary users, often driven by business considerations. Consistent with this observation, none of the reviewed apps provided settings for visiting bystanders or offered consent prompts when others, such as domestic workers~\cite{he2025exploring}, might be captured by sensors or included in shared data streams.

Our findings also resonate with prior work on technology “non-use,” which frames non-use not as a simple absence of interaction but as a meaningful socio-technical practice shaped by power, context, and culture~\cite{satchell2009beyond,baumer2015importance}. In our study, bystanders were often non-users of smart home apps yet remained deeply implicated in their data practices. Because policies and interfaces typically define users as account holders or device owners, bystanders lack mechanisms to refuse, negotiate, or opt out, effectively stripping them of agency in favor of owners and vendors. Many of Chinese smart home apps we analyzed collected highly identifiable bystander data in opaque and unexpected ways while failing to distinguish between primary users and bystanders. This flattening of roles ignores differing expectations around data retention, sharing, and deletion~\cite{barkhuus2012mismeasurement,mcmillan2013categorised}. We argue that bystander non-use, together with bystanders’ privacy expectations and concerns, should be treated as a core design consideration in smart home apps and related laws and regulations, rather than as a peripheral edge case.%%

\subsection{Inconsistencies Between Privacy Policies and UX/UI Design Practices}

In response to RQ3, our traceability analysis revealed a pronounced gap between the privacy practices promised in Chinese smart home apps’ policies and what users could actually observe or control through their UX/UI designs. While many apps formally declared adherence to China’s PIPL and CSL, these commitments often lacked tangible UI support (see \S\ref{traceability_analyse}). For example, 29/49 apps failed to provide clear in-app privacy controls (e.g., permission management or personalized content settings) despite referencing such mechanisms in their policies, resulting in a \textit{Partial} traceability rating. Similarly, although 28 apps mentioned data deletion or retention mechanisms in their policy texts, they lacked visible or functional options within their settings. As discussed in \S\ref{UX_privacy_controls}, some privacy settings were buried under multiple layers of menus or presented without explanatory labels, limiting users’ ability to make informed choices. This disconnect was particularly evident for opt-out mechanisms related to personalized content: in four apps, these controls were entirely absent, resulting in a \textit{Broken} traceability classification.

Beyond policy-to-UI mismatches, we also observed inconsistencies between privacy labels and actual data practices (see \S\ref{result_privacy_label}). For instance, 26 apps indicated in their privacy labels that they did not collect \textit{Data Used to Track You}, despite simultaneously disclosing the use of third-party SDKs (e.g., for analytics or push notifications) in their privacy policies and UI flows (RQ1, RQ2, RQ3). Consistent with prior work~\cite{koch2022keeping,kollnig2022goodbye,li2022understanding}, our findings suggest that the accuracy of privacy labels is not systematically verified during the app approval process in China’s App Store. This raises concerns about potential mislabeling and suggests that privacy labeling remains more symbolic than reliable~\cite{zhang2022usable,khandelwal2023comparing}. More broadly, the systemic nature of these mismatches reflects a tendency to treat privacy as a legal documentation task rather than a core product requirement. Although all 49 apps acknowledged their legal obligation under the CSL (Art.~35) to share user data with public security authorities, none provided visual indicators or consent flows explaining how or when such access might occur (see Table~\ref{tab:traceability_table}). Instead, data control and deletion options were often accompanied by vague exceptions such as “as permitted by law” or “unless required by national interest.” These omissions both hinder user understanding and obscure critical conditions, thereby undermining users’ perceived control over their personal data.%%

\subsection{Recommendations to Consider and Design for Smart Home Bystander Privacy}

Our findings addressing RQ1, RQ2, and RQ3 reveal a pervasive lack of consideration for bystander privacy across smart home apps, regardless of device type. In response, we propose a set of concrete recommendations aimed at improving transparency, privacy control, and protections for bystanders within smart home apps and ecosystems. These recommendations are grounded in our empirical findings and directly address the systemic neglect of bystander perspectives identified in our UX/UI analysis.

\textbf{Make bystander data collection and processing transparent during onboarding.} Smart home apps should inform device owners or primary users during account creation about potential data collection and processing involving bystanders and non-users, such as guests, cohabitants, or domestic workers (RQ1). This includes clearly communicating that certain device features (e.g., cameras or voice assistants) may inadvertently capture bystander data. The onboarding process should include concise, user-friendly disclosures and recommend responsible usage practices. Additionally, apps could provide template reminders or notifications encouraging account holders to delete bystander data after gatherings or periods of shared use~\cite{bernd2022balancing,geeng2019s,slupska2022they,h2022monitoring,wang2022individual,tabassum2023exploring,zeng2019understanding,despres2024my}.

\textbf{Integrate dedicated bystander privacy controls.} Our analysis showed that privacy controls for bystanders were often absent or embedded within broader family management features (RQ2, RQ3). To better support bystanders in multi-user smart home environments, apps should provide dedicated privacy settings that allow device owners or primary users to manage bystander-related data practices explicitly. Such controls could enable users to toggle data collection for specific bystander categories (e.g., children or visitors), apply data minimization techniques (e.g., face blurring in video streams), or temporarily suspend data collection during known visiting periods (e.g., via privacy zones). Apps should also offer real-time prompts or notifications to encourage timely adjustment of these settings.

\textbf{Offer consent-oriented guest modes.} Although apps controlling door locks commonly support guest access management (see \S\ref{UX_privacy_controls}; RQ2), these features primarily focus on authentication rather than data protection. Building on prior proposals for guest modes (e.g.,~\cite{geeng2019s,marky2022you,lau2018alexa,yao2019defending,alshehri2023exploring,albayaydh2023examining}), we recommend extending guest modes to include: (1) explicit opt-in consent prompts for data collection; (2) options for assigning limited data management responsibilities to bystanders (e.g., domestic workers); and (3) app-based or physical interfaces that clearly communicate active privacy settings to all individuals present.

\textbf{Enable bystander-initiated data deletion or objection.} Currently, data deletion mechanisms are exclusively tied to the primary user account (RQ1, RQ2, RQ3), leaving bystanders without meaningful control over their personal data. We recommend that smart home apps introduce mechanisms that allow bystanders to request data deletion or object to data processing, either directly through secure portals or workflow notifications, or indirectly via the primary user. Such mechanisms could support negotiation between bystanders (e.g., domestic workers) and device owners (e.g., employers) regarding recorded data and consent~\cite{he2025exploring}.%%

\subsection{Design Recommendations to Improve Privacy-focused UI Features}

In response to the UI shortcomings identified when addressing RQ2 and the policy-to-practice gaps found (RQ3), we recommend that smart home device developers integrate transparent, accessible, and actionable privacy controls directly into the app UI. Our traceability analysis (\S\ref{traceability_analyse}) shows that regulatory compliance alone is insufficient---users require clear, contextual cues and meaningful interaction points to exercise agency over their data. Although many Chinese users have normalized state surveillance practices~\cite{he2025privacy,he2025exploring}, developers should still surface critical visual indicators, particularly notifications disclosing when personal data is shared with the state for national security purposes. To address state-mandated data sharing practices, apps should provide proactive notices or adopt explainable privacy design patterns clarifying when and how such disclosures may occur. Even when user consent is not legally required, transparency around access conditions, such as legal triggers and oversight mechanisms, can enhance perceived control and align with international best practices (e.g., Art.~13 of GDPR~\cite{gdprart13}). Applying principles of layered notice and progressive disclosure can further improve user understanding and engagement with privacy settings, such as incorporating just-in-time notices that explain data use in context~\cite{kitkowska2020enhancing}.

\subsection{Design Recommendations to Improve Privacy Labels}

To address communication failures in privacy policies (RQ1) and UIs (RQ2), as well as the inconsistencies between them identified (RQ3), our analysis of Apple App Store privacy labels revealed that these labels in Chinese smart home apps are largely symbolic and unverified. App Store privacy labels are self-reported by developers, and the information is not verified during app approval~\cite{koch2022keeping}, contributing to systematic underreporting (see \S\ref{result_privacy_label}). Building on prior work~\cite{emami2021privacy,emami2020ask}, our findings reinforce the need for a consistent, transparent, and user-centric approach to privacy disclosure---particularly in the context of Chinese smart home apps on iOS. Privacy labels must move beyond symbolic compliance to become actionable and trustworthy mechanisms that genuinely support user awareness and control.%%

\subsubsection{Recommendations for App Developers}

Smart home apps, often interconnected with a range of hardware devices, should adopt layered privacy labels that present core attributes (e.g., data types collected, data retention periods, third-party sharing) in a glanceable summary, with expandable sections offering more technical and legal detail. This approach aligns with Emami-Naeini et al.'s findings that users prefer concise overviews with optional in-depth information~\cite{emami2020ask}. In the Chinese context, such labels can also clarify vague policy language commonly used in apps, such as ``data may be shared as required by relevant laws for national security purposes.'' Effective privacy communication also requires localization to ensure cultural and linguistic relevance. Developers should use plain-language explanations and localized visual cues to enhance user comprehension, especially among non-expert users~\cite{he2025privacy}.

While Chinese regulations are often broadly worded, their open-ended nature allows for adaptable enforcement~\cite{cao2016chinese}. Developers need to translate these ambiguous legal requirements into practical design guidelines for privacy label visualizations. Regulatory bodies such as the Ministry of Industry and Information Technology (MIIT) and the CAC could support this process by issuing standardized guidelines for the smart home sector, thereby reducing inconsistencies across providers~\cite{habib2019empirical}. For example, authorities could mandate the visibility of key privacy rights, such as consent withdrawal, data retention periods, and access logs. Standardizing these elements would help ensure that users can consistently locate and understand privacy controls, regardless of the provider.%%%

\subsubsection{Recommendations for App Platforms}

Given the widespread use of third-party SDKs, developers must ensure that data-sharing disclosures are both specific and verifiable. Current inconsistencies between privacy labels and actual SDK behavior undermine user trust~\cite{koch2022keeping}. Platforms like the Apple App Store should enforce cross-verification between privacy labels, privacy policies, and app behavior, potentially through automated or crowdsourced audits, as proposed by prior studies~\cite{libert2018automated,wilson2018analyzing}. Additionally, the Apple App Store could require all IoT and smart home app developers to include enhanced privacy labels or visual diagrams detailing data handling practices, particularly for sensitive sensors, such as cameras and microphones~\cite{chen2024trustmark}.
%\subsection{Suggestions to Policymaker}
%\textcolor{blue}{Our findings reveal the practical limitations of a one-size-fits-all regulatory framework (PIPL) when applied to the heterogeneous smart home ecosystem. While the PIPL does differentiate Sensitive PII, such as biometrics and health data, and requires separate consent for its processing (PIPL, Art. 28-29), our analysis shows that this distinction is poorly implemented in practice---a gap that is even more pronounced when considering bystander privacy. We argue for future regulatory guidance to adopt a more explicit, risk-based approach tailored to the smart home sector, one that explicitly considers the privacy needs of bystanders. This could resemble the tiered ``high-risk'' application classifications seen in other modern regulations, such as the EU's AI Act~\cite{golpayegani2023high} and multi-level system for privacy risk management~\cite{he2025privacy}. Such a framework would provide clearer, device-specific compliance obligations, distinguishing between high-risk devices (e.g., cameras, health monitors) and low-risk devices (e.g., lights, blenders). This would not only provide stronger protections where they are needed most---as highlighted by the bystander privacy concerns with high-risk devices---but also clarify the compliance burden for low-risk innovators, addressing the ambiguities that our study suggests are leading to widespread systematic neglect.}%%

\subsection{Limitations and Future Work} \label{limitation}

Our analysis has several limitations that could have affected the generalizability of our findings. First, our study focused on a single app ecosystem, limiting direct applicability to other platforms (e.g., Android) and to different cultural or regulatory frameworks. Nevertheless, we carefully documented our codebook---developed and iteratively refined during the privacy policy analysis---as well as our UX/UI analysis tasks, providing a transparent foundation for replication and extension. Future studies could build on these materials to examine a broader, cross-platform sample of smart home apps, enabling comparative analyses across ecosystems and regulatory frameworks.

Second, our sample comprised heterogeneous smart home apps (see Fig.~\ref{fig:app_distribution}), ranging from single-sensor devices to ecosystem-level controllers. While this diversity supported our goal of examining cross-cutting compliance with PIPL, which applies across app categories, different application domains (e.g., financial, audiovisual, or health) and device modalities may entail distinct risk profiles and appropriate privacy controls, particularly for bystanders. For example, the legal requirements and bystander privacy implications of a camera app processing biometric data may differ substantially from those of a smart blender. Future work could therefore adopt modality-sensitive and stratified research designs to provide more fine-grained insights.

Third, our analysis was limited to publicly available privacy policies and observable UX/UI features and practices. We were unable to verify backend data-handling operations or assess the accuracy and completeness of policy disclosures beyond what is visible to users. It is possible that companies implement additional privacy safeguards or data-handling practices that are not explicitly documented or surfaced in the interface~\cite{he2025privacy}. Future research could incorporate technical audits, backend data flow analyses, or collaborative studies with industry partners to more rigorously assess alignment between documented policies and actual data-processing practices.

Finally, given the dynamic regulatory landscape in China, particularly with respect to personal data protection and national security requirements, our findings represent a snapshot in time. As privacy policies, app functionalities, and data practices continue to evolve in response to regulatory changes and technological developments, longitudinal and ongoing analyses will be necessary to capture emerging trends and shifts in privacy practices and usability standards within the smart home domain.

\section{Conclusion}
In this study, we examined the privacy practices of 49 Chinese smart home apps, revealing a significant mismatch between documented privacy policies and actual UX/UI features. Through a comprehensive analysis of app privacy policies, interface features, and Apple App Store privacy labels, complemented by traceability analysis, we uncovered widespread inconsistencies and critical gaps. While formal compliance with the PIPL and CSL was common, most apps failed to provide meaningful control over personal data---particularly for non-primary users such as bystanders. We also identified systemic design shortcomings, including opaque consent mechanisms, inadequate deletion pathways, and limited agency for bystanders such as visitors or domestic workers. Our findings demonstrate that legal compliance alone is insufficient to ensure robust privacy protections in complex, multi-user smart home environments. By highlighting these overlooked vulnerabilities, we advocate for a more inclusive, transparent, and context-sensitive approach to smart home privacy---one that accounts for shared domestic realities and extends protections beyond the primary user. Furthermore, we offer concrete design and policy recommendations to support the development of bystander-aware smart home ecosystems that prioritize privacy for all affected individuals.

%% consistent spelling of the heading%

\begin{acks}
We thank the anonymous CHI reviewers for their valuable feedback. ChatGPT was used to assist with proofreading and grammar checking. This work was partially supported by the Generalitat Valenciana through the PROMETEO CIPROM/2023/23 project and by the Spanish Government under grant PID2023-151536OB-I00.
\end{acks}
% The next two lines define the bibliography style to be used, and
%% the bibliography file.%
\bibliographystyle{ACM-Reference-Format}
\bibliography{reference}

\appendix
%\section{Chinese Smart Home APPs}
%\input{content/table/app}
\onecolumn
\section{UI Testing}
\begin{table*}[ht!]
\centering
\fontsize{6pt}{6pt}\selectfont
\resizebox{\textwidth}{!}{%
\begin{tabular}{ll}
\toprule
\textbf{Tasks for Bystanders} & \textbf{Description} \\
\midrule
Configuring Privacy Settings for Bystanders &
\makecell[tl]{• Navigating settings related to bystander privacy.\\[0pt]
• Enabling features such as ``Guest Mode'' or ``Privacy Mode''.\\[0pt]
• Setting automatic rules to disable cameras and/or microphones when guests \\ are present.} \\
%\midrule
Managing Data Collection and Sharing Preferences Affecting Bystanders &
\makecell[tl]{• Adjusting settings to limit data collection when bystanders are present.\\[0pt]
• Configuring permissions for sharing data that may involve bystanders' information \\ (e.g., images, audio recordings).} \\
%\midrule
Informing Bystanders About Device Status &
\makecell[tl]{• Ensuring visual or auditory indicators are active when devices are recording.\\[0pt]
• Customizing notifications or alerts to inform bystanders about device usage.} \\
%\midrule
Sharing Access or Control with Bystanders &
\makecell[tl]{• Setting up guest accounts or temporary access for bystanders.\\[0pt]
• Allowing bystanders to disable devices or modify privacy settings.} \\
%\midrule
Reviewing and Deleting Data Involving Bystanders &
\makecell[tl]{• Accessing stored data that may include bystanders (e.g., video clips, audio \\ recordings).\\[0pt]
• Deleting or anonymizing data that involves bystanders.} \\
\bottomrule
\end{tabular}}

\vspace{0cm}

\resizebox{1\textwidth}{!}{%
\begin{tabular}{ll}
\toprule
\textbf{General Privacy Tasks} & \textbf{Description} \\
\midrule
Accessing and Understanding the Privacy Policy &
\makecell[tl]{• Locating the privacy policy within the app.\\[0pt]
• Reading and comprehending the information provided.} \\
%\midrule
Configuring Privacy and Security Settings &
\makecell[tl]{• Navigating the settings menu.\\[0pt]
• Adjusting permissions for data collection, location access, microphone use, etc.} \\
%\midrule
Managing Data Collection and Sharing Preferences &
\makecell[tl]{• Opting in or out of data collection features.\\[0pt]
• Modifying settings related to sharing data with third parties.} \\
%\midrule
Deleting Personal Data or Deactivating the Account &
\makecell[tl]{• Finding options to delete personal data.\\[0pt]
• Understanding the consequences of data deletion.\\[0pt]
• Initiating account deactivation or deletion.} \\
\bottomrule
\end{tabular}}

\caption{Tasks related to configuring privacy settings for bystanders ($n=5$) and general privacy tasks ($n=4$).}
\label{tab:privacy_tasks}
\end{table*}%%

\section{Evaluation Methods}
\begin{table*}[ht!]
\centering
\fontsize{6pt}{6pt}\selectfont
\resizebox{\textwidth}{!}{%
\begin{tabular}{ll}
\toprule
\textbf{Heuristic} & \textbf{Description} \\
\midrule
Visibility of System Status & Smart devices should clearly indicate when they are recording or transmitting data, ensuring bystanders are aware. \\
%\hline
Match Between System and Real World & Labels such as ``AI Guardian'' or ``Cloud Scene'' should use familiar language like ``Motion Alert'' or ``Camera On.'' \\
%\hline
User Control and Freedom & Users should be able to undo settings or temporarily disable monitoring for specific zones or individuals. \\
%\hline
Consistency and Standards & Privacy options should maintain consistent layout, naming, and iconography across all devices and menus. \\
%\hline
Error Prevention & The system should alert users before enabling functions that could impact others' privacy. \\
%\hline
Recognition Over Recall & Privacy tools should be visible and easily recognizable rather than requiring users to remember deep navigation paths. \\
%\hline
Flexibility and Efficiency of Use & Experienced users should be able to create reusable privacy presets for common scenarios, such as guest visits. \\
%\hline
Aesthetic and Minimalist Design & Avoid cluttered settings; key privacy controls should be prominently displayed and easy to access. \\
%\hline
Recognize, Diagnose, and Recover from Errors & Systems should notify users of privacy setting failures and provide guidance to restore secure defaults. \\
%\hline
Help and Documentation & Provide in-app tutorials and localized help resources to guide users in protecting bystander privacy. \\
\bottomrule
\end{tabular}}
\caption{Adapted usability heuristics for privacy in Chinese smart home apps.}
\label{tab:heuristics}
\end{table*}

\vspace{0cm}

\begin{table*}[ht!]
\centering
\fontsize{6pt}{6pt}\selectfont
\resizebox{\textwidth}{!}{%
\begin{tabular}{p{3cm}p{10cm}}
\toprule
\textbf{Guideline} & \textbf{Description} \\
\midrule
Data Minimization & Apps should only collect information required for core functionality. Bystanders should not need to provide ID or voice data unless essential. \\
%\hline
Anonymization & Where possible, systems should blur or anonymize bystanders' faces or voices, particularly in shared or guest contexts. \\
%\hline
User Control & Users must have clear options to consent, manage, or opt out of data practices—for both themselves and others in the household. \\
%\hline
Privacy Settings & Privacy controls should be easy to locate and include options specific to live-in, visiting, or uninvolved bystanders. \\
%\hline
Right to be Forgotten & Smart home systems should allow deletion of voice logs, video clips, and facial data, particularly for one-time guests. \\
%\hline
Access and Correctness & Users should be able to view, verify, and correct stored data, including any data related to bystanders. \\
%\hline
Functionality & Apps should provide basic services without requiring unnecessary permissions or full access to cameras and microphones. \\
%\hline
Transparency & Data collection and sharing practices must be clearly explained, including retention periods and access rights. \\
\bottomrule
\end{tabular}}
\caption{Adapted Privacy-by-Design principles for Chinese smart home apps.}
\label{tab:pbdd}
\end{table*}%%%%%%

\onecolumn
\section{Themes and Corresponding Codes}
\begin{table*}[h]
\centering
%\fontsize{8pt}{10pt}\selectfont
\footnotesize
\resizebox{1\textwidth}{!}{%
\begin{tabular}{llp{8cm}}
\toprule
\textbf{Themes} & \textbf{Code Name} & \textbf{Code Description} \\
\midrule
\multirow{7}{*}{Policy Scope and Version Information} 
& Policy Scope (Products/Services) & Specifies that the policy covers hardware, companion apps, cloud services, and websites. \\
& Effective/Update Dates & States the latest update and effective dates, and notes whether they coincide. \\
& Change Notification Mechanism & Commits to notifying users of material changes via pop-ups, email, or announcements. \\
& Change Explainability & Clearly describes the types of changes (e.g., new features or data processing purposes). \\
& Access to Historical Versions & Provides links to prior versions or instructions for obtaining them. \\
& Dedicated Privacy Site & Offers a standalone privacy center or website for versions, requests, and complaints. \\
& Continued Use Implies Consent & States that continued use constitutes acceptance of the revised policy. \\
\hline
\multirow{6}{*}{Readability and Information Presentation} 
& Length \& Reading Burden & Notes the document length and estimated reading time for usability assessment. \\
& Structural Organization & Uses sections, subsections, and numbered items for easier navigation. \\
& Key Highlighting & Highlights critical items using bold text or color. \\
& Supportive Links & Provides links to SDK lists, permission purposes, and glossaries. \\
& Visual Aids & Uses tables or diagrams to illustrate permissions and scenarios. \\
& Plain-Language Summary & Provides a simplified summary for lay users. \\
\hline
\multirow{9}{*}{Data Collection Methods (Direct/Indirect)} 
& Direct Collection – Registration \& Real-Name & User-entered data for account setup, such as phone number, verification code, and password. \\
& Third-Party Login Data & Avatar, nickname, region, etc., obtained from WeChat, Alipay, or Apple ID with consent. \\
& Direct Collection – Optional Profile & User-supplied optional fields, such as username, avatar, region, or gender. \\
& Indirect Collection – Device Identifiers & IMEI, OAID, MAC address, model, OS, timezone, and similar identifiers. \\
& Indirect Collection – Network \& Logs & IP address, Wi-Fi SSID/BSSID, crash logs, interaction logs, and anomaly reports. \\
& Indirect Collection – Location & Precise or approximate location (GPS/IP/cell) for onboarding or geofencing. \\
& Permissions \& Multimedia & Use of camera, photos, microphone, Bluetooth, and instructions for disabling them. \\
& Device \& Scenario Data & Device telemetry, energy consumption, automation scripts, cleaning routes, and alerts. \\
& AI/Personalization Data & Data generated from historical environment, energy usage, temperature, and humidity for personalized services. \\
\hline
\multirow{4}{*}{Sensitive Information and Special Categories} 
& Payment \& Financial Data & Order and payment information processed only during transactions. \\
& Biometric Data & Face, fingerprint, voiceprint, vein patterns, etc. \\
& Health \& Physiological Data & Heart rate, sleep, blood pressure, blood sugar, menstrual cycle, weight, etc. \\
& Children's Data (Dedicated Section) & Defines minors, guardian consent, and special handling rules. \\
\hline
\multirow{5}{*}{Data Storage and Cross-Border Transfers}
& Onshore-Only Storage (China) & All data stored domestically by default; no routine cross-border transfers. \\
& Conditional Cross-Border Transfer & Data stored domestically by default; outbound transfers only with assessment, contracts, and separate consent. \\
& Global/Regional Deployment & Multi-region cloud deployments with SCC/GDPR-aligned safeguards. \\
& Relay-Only, No Persistent Storage & Real-time audio/video relayed abroad without storage. \\
& Retention \& Data Minimization & Specifies minimal retention periods and automatic deletion schedules. \\
\hline
\multirow{5}{*}{Data Sharing and Disclosure Recipients}
& Household/Shared Use Data Sharing & Mechanisms for sharing devices or data with family members. \\
& Impact on Secondary Users and Bystanders & Bystander or visitor capture, notification, and consent processes. \\
& Third-Party Services \& SDKs & Categories and purposes of sharing with payment, cloud, advertising, or content partners. \\
& Ads \& Profiling Controls & Ability to opt out of personalization and instructions to exercise rights. \\
& Lawful Disclosure & Disclosures to public or national security authorities under law; consent exemptions noted. \\
\hline
\multirow{8}{*}{Data Control and User Rights}
& Account Deletion & Self-service deletion paths; immediate or with cooling-off periods. \\
& Device Unbind \& Delete & Unbinding a device deletes its associated data. \\
& Automatic Purge \& Log Rotation & Periodic deletion policies for logs and media. \\
& Legal Constraints on Deletion & Deletion may be delayed or denied during investigations. \\
& Data Export/Portability & Specifies what data can be exported and how (e.g., email or self-service). \\
& Children's Rights & Rights exercisable by children or guardians (access, rectify, delete). \\
& Bystander Rights & Channels for removal or masking of bystanders’ data and responsibility allocation. \\
& Deceased Users' Data & Rights of close relatives and limits for legitimate interests. \\
\hline
\multirow{7}{*}{Information Security and Management}
& Encryption \& Transport Security & TLS/SSL, at-rest encryption, hashing, and de-identification. \\
& MLPS/National Standards Compliance & Declares compliance with MLPS 2.0 and GB/T standards. \\
& International Certifications & ISO/IEC 27001/27701, SOC reports, and similar certifications. \\
& Access Control \& Least Privilege & Implements least privilege, audit logs, and periodic reviews. \\
& Incident Response \& Breach Notice & Provides playbooks, handling flows, and notification timelines. \\
& Third-Party Due Diligence & Security reviews and contractual safeguards for partners. \\
& Liability \& Disclaimer & Acknowledges Internet risks and shared responsibility with users. \\
\bottomrule
\end{tabular}}
\caption{Themes and corresponding codes for the analysis of smart home app privacy policies ($n=49$).}
\label{tab:codebook-en}
\end{table*}

\onecolumn
\section{Privacy Labels}
\subsection*{Data Used to Track You}

\begin{center}
\fontsize{5pt}{4pt}\selectfont
\begin{tabular}{l|ccc}
\hline
\textbf{ID} & Identifiers & Usage Data & Diagnostics \\
\hline
APP2 & \textbf{*} &  &  \\
APP4 &  & \textbf{*} &  \\
APP5 &  & \textbf{*} &  \\
APP6 & \textbf{*} &  &  \\
APP8 & \textbf{*} & \textbf{*} &  \\
APP10 & \textbf{*} &  &  \\
APP11 & \textbf{*} &  &  \\
APP13 & \textbf{*} &  &  \\
APP16 & \textbf{*} &  &  \\
APP17 & \textbf{*} &  &  \\
APP20 & \textbf{*} &  & \textbf{*} \\
APP24 & \textbf{*} & \textbf{*} &  \\
APP25 & \textbf{*} &  &  \\
APP27 & \textbf{*} &  &  \\
APP29 & \textbf{*} & \textbf{*} &  \\
APP30 & \textbf{*} &  &  \\
APP31 & \textbf{*} & \textbf{*} &  \\
APP32 & \textbf{*} & \textbf{*} &  \\
APP34 & \textbf{*} &  &  \\
APP35 & \textbf{*} &  &  \\
APP37 & \textbf{*} &  &  \\
APP41 &  &  & \textbf{*} \\
APP44 & \textbf{*} & \textbf{*} &  \\
\hline
\end{tabular}
\end{center}

\subsection*{Data Linked to You}
\begin{center}
\fontsize{5pt}{4pt}\selectfont
\begin{tabular}{l|cccccccc}
\hline
\textbf{ID} & Contact Info & Contacts & User Content & Search History & Purchases & Financial Info & Health & Location \\
\hline
APP1 & \textbf{*} & & \textbf{*} & & & & & \\
APP2 & \textbf{*} & & \textbf{*} & & & & & \textbf{*} \\
APP4 & \textbf{*} & & & & & & & \\
APP5 & & \textbf{*} & \textbf{*} & & & & & \\
APP6 & \textbf{*} & & \textbf{*} & \textbf{*} & & & & \\
APP8 & \textbf{*} & \textbf{*} & \textbf{*} & \textbf{*} & & & & \\
APP9 & \textbf{*} & \textbf{*} & \textbf{*} & & & & \textbf{*} & \textbf{*} \\
APP11 & \textbf{*} & \textbf{*} & & \textbf{*} & \textbf{*} & \textbf{*} & & \\
APP12 & \textbf{*} & & & \textbf{*} & & & & \\
APP16 & \textbf{*} & \textbf{*} & \textbf{*} & & & & \textbf{*} & \textbf{*} \\
APP18 & \textbf{*} & & \textbf{*} & & \textbf{*} & & & \\
APP19 & \textbf{*} & & & & & & & \\
APP21 & \textbf{*} & & \textbf{*} & & & & \textbf{*} & \textbf{*} \\
APP23 & \textbf{*} & & & & & & & \\
APP24 & \textbf{*} & & \textbf{*} & & & & & \\
APP25 & \textbf{*} & & & & & & & \\
APP26 & \textbf{*} & & \textbf{*} & & & & & \textbf{*} \\
APP27 & \textbf{*} & & \textbf{*} & \textbf{*} & & & & \\
APP29 & & & & & & & & \textbf{*} \\
APP30 & \textbf{*} & & \textbf{*} & \textbf{*} & \textbf{*} & \textbf{*} & & \textbf{*} \\
APP31 & \textbf{*} & & \textbf{*} & & & & & \\
APP32 & & & \textbf{*} & & & & & \\
APP34 & \textbf{*} & & & & & & & \\
APP36 & \textbf{*} & & \textbf{*} & & & & & \\
APP39 & \textbf{*} & & \textbf{*} & & & & & \textbf{*} \\
APP42 & \textbf{*} & & & & & & & \textbf{*} \\
APP44 & \textbf{*} & \textbf{*} & \textbf{*} & & & & & \\
APP47 & \textbf{*} & & \textbf{*} & & & & & \\
APP49 & \textbf{*} & & & & & & \textbf{*} & \textbf{*} \\
\hline
\end{tabular}
\end{center}

\subsection*{Data Not Used to Track You}
\begin{table*}[h]
\centering
\fontsize{5pt}{4pt}\selectfont
\begin{tabular}{l|cccccccccc}
\hline
\textbf{ID} & Location & Contact Info & Contacts & User Content & Identifiers & Usage Data & Diagnostics & Search History & Purchases & Other Data \\
\hline
APP1 & \textbf{*} & \textbf{*} & & \textbf{*} & & \textbf{*} & \textbf{*} & & & \textbf{*} \\
APP2 & & & & & & & \textbf{*} & & & \\
APP3 & & & & & & & \textbf{*} & & & \\
APP4 & \textbf{*} & \textbf{*} & & \textbf{*} & \textbf{*} & & \textbf{*} & & & \\
APP5 & \textbf{*} & & & \textbf{*} & & & \textbf{*} & & & \\
APP6 & \textbf{*} & & & & & & & & & \\
APP8 & \textbf{*} & & & & & & & & & \textbf{*} \\
APP9 & & & & & & \textbf{*} & \textbf{*} & & & \textbf{*} \\
APP11 & \textbf{*} & & & \textbf{*} & & \textbf{*} & \textbf{*} & & & \\
APP12 & \textbf{*} & & & \textbf{*} & & \textbf{*} & \textbf{*} & & & \\
APP13 & \textbf{*} & \textbf{*} & \textbf{*} & \textbf{*} & \textbf{*} & \textbf{*} & \textbf{*} & \textbf{*} & \textbf{*} & \textbf{*} \\
APP14 & \textbf{*} & & & \textbf{*} & \textbf{*} & & \textbf{*} & & & \\
APP15 & & \textbf{*} & & & & \textbf{*} & \textbf{*} & & & \\
APP16 & \textbf{*} & & & \textbf{*} & \textbf{*} & \textbf{*} & \textbf{*} & & & \\
APP17 & \textbf{*} & \textbf{*} & & \textbf{*} & & \textbf{*} & \textbf{*} & & & \\
APP18 & \textbf{*} & \textbf{*} & & \textbf{*} & & & & & & \\
APP19 & \textbf{*} & \textbf{*} & & \textbf{*} & & \textbf{*} & \textbf{*} & & & \\
APP20 & \textbf{*} & & & & \textbf{*} & & \textbf{*} & & & \\
APP21 & & & & \textbf{*} & & & \textbf{*} & & & \\
APP22 & \textbf{*} & \textbf{*} & & & & & & & & \\
APP23 & \textbf{*} & & & \textbf{*} & & \textbf{*} & & & & \\
APP24 & \textbf{*} & & & & & & \textbf{*} & & & \\
APP25 & \textbf{*} & & \textbf{*} & \textbf{*} & & \textbf{*} & \textbf{*} & & & \\
APP26 & & & & & & & \textbf{*} & & & \\
APP27 & \textbf{*} & & & \textbf{*} & \textbf{*} & \textbf{*} & \textbf{*} & & & \textbf{*} \\
APP31 & \textbf{*} & & & & \textbf{*} & \textbf{*} & \textbf{*} & & & \\
APP32 & \textbf{*} & & & & \textbf{*} & & & & & \\
APP34 & \textbf{*} & \textbf{*} & & & \textbf{*} & \textbf{*} & \textbf{*} & & & \\
APP35 & \textbf{*} & \textbf{*} & & \textbf{*} & & & \textbf{*} & & & \\
APP36 & \textbf{*} & & & & \textbf{*} & \textbf{*} & \textbf{*} & & & \\
APP37 & \textbf{*} & \textbf{*} & & \textbf{*} & \textbf{*} & \textbf{*} & \textbf{*} & \textbf{*} & & \textbf{*} \\
APP40 & \textbf{*} & \textbf{*} & & & & \textbf{*} & & & & \\
APP41 & \textbf{*} & \textbf{*} & & \textbf{*} & & \textbf{*} & \textbf{*} & & & \\
APP42 & & & & & & \textbf{*} & \textbf{*} & & & \\
APP43 & \textbf{*} & \textbf{*} & & \textbf{*} & \textbf{*} & \textbf{*} & \textbf{*} & & & \textbf{*} \\
APP44 & \textbf{*} & & & & & & \textbf{*} & & & \\
APP46 & \textbf{*} & & & & \textbf{*} & & & & & \textbf{*} \\
APP47 & \textbf{*} & & & & & \textbf{*} & \textbf{*} & & & \\
APP49 & & & & & \textbf{*} & & & & & \\
\hline
\end{tabular}
\caption{Privacy labels for 49 selected Chinese smart home devices from the Apple App Store in the Mainland China market (as of January 2025). The three tables exclude blank labels for apps that did not provide this information.}
\label{tab:privacy_labels}
\end{table*}%%%

%\end{CJK*}
\end{document}